\documentclass[aip,jcp,preprint,citeautoscript,floatfix]{revtex4-1}

\usepackage{graphicx}
\usepackage{amsmath}
\usepackage{amsfonts}
\usepackage{amssymb}
\usepackage{amsbsy}
\usepackage{multirow}
\usepackage{bm}
\usepackage{color}

\usepackage{epstopdf}
\usepackage{ulem}
\usepackage{verbatim}

\usepackage{placeins}

\usepackage{calligra}
\usepackage{calrsfs}
\usepackage[mathscr]{euscript}

\usepackage{xr}

\DeclareMathAlphabet\mathbfcal{OMS}{cmsy}{b}{n}

\usepackage[english]{babel}

\newcommand{\bt}{\beta}
\newcommand{\dt}{\delta}
\newcommand{\Dt}{\Delta}
\newcommand{\la}{\lambda}
\newcommand{\ga}{\gamma}
\newcommand{\Ga}{\Gamma}
\newcommand{\kT}{k_B T}

\newcommand{\llangle}{\left\langle}
\newcommand{\rrangle}{\right\rangle}

\newcommand{\osm}{solute}
\newcommand{\conc}{concentration}

\newcommand{\kbt}{k_BT}
\newcommand{\omu}{\overbar{\mu}}

\newcommand{\wt}[1]{\widetilde{{#1}}}

\newcommand{\ve}{\varepsilon}
\newcommand{\ove}{\overline{\varepsilon}}
\newcommand{\rdbracket}[1]{\left( {#1} \right)}
\newcommand{\sqbracket}[1]{\left[ {#1} \right]}

\newcommand{\overbar}[1]{\mkern 1.5mu\overline{\mkern-1.5mu#1\mkern-1.5mu}\mkern 1.5mu}

\newcommand{\half}{\frac{1}{2}}
\usepackage[font=footnotesize,labelfont=bf]{caption}

\newcommand{\op}{{o'}}
\newcommand{\oop}{{o\op}}

\newcommand{\Nw}{N_w}
\newcommand{\nw}{n_w}
\newcommand{\mo}{{m_o}}
\newcommand{\mop}{{m_\op}}
\newcommand{\no}{{n_o}}
\newcommand{\No}{{N_o}}
\newcommand{\Nop}{N_\op}
\newcommand{\xo}{x_o}
\newcommand{\xw}{x_w}
\newcommand{\xop}{x_\op}
\newcommand{\setno}{\{\no\}}
\newcommand{\setNo}{\{\No\}}
\newcommand{\setmuo}{\{\muo\}}
\newcommand{\sumo}{\sum_o}
\newcommand{\sumno}{\sum_{\{n_o \ge 0\}}}
\newcommand{\sumNo}{\sum_{\{N_o \ge 0\}}}
\newcommand{\hmuo}{\hat\mu_o}
\newcommand{\muo}{\mu_o}
\newcommand{\hmuw}{\hat\mu_w}
\newcommand{\muw}{\mu_w}
\newcommand{\muwo}{\mu_{w}^\phi}
\newcommand{\muost}{\muo^*}

\newcommand{\hgamma}{\hat{\gamma}}

\newcommand{\lo}{\la_o}
\newcommand{\ao}{a_o}
\newcommand{\aop}{a_\op}
\newcommand{\ato}{a_{t o}}
\renewcommand{\atop}{a_{t \op}}		% highjacking \atop command
\newcommand{\ko}{k_o}
\newcommand{\kop}{k_\op}
\newcommand{\eti}{\eta_i}

\newcommand{\sumop}{\sum_\op}
\newcommand{\sumoop}{\sum_{o,o'}}
\newcommand{\veoop}{\ve_{o\op}}
\newcommand{\veoo}{\ve_{oo}}

\newcommand{\dd}{{\rm d}}
\newcommand{\ptl}{\partial}
\newcommand{\inv}{^{-1}}

\newcommand{\dM}{\dd M}
\newcommand{\dno}{\dd \no}
\newcommand{\dNo}{\dd \No}
\newcommand{\dmuo}{\dd \muo}
\newcommand{\dmuw}{\dd \muw}
\newcommand{\dtmuw}{\dt \muw}
\newcommand{\dga}{\dd \ga}

\newcommand{\Gao}{\Ga_o}
\newcommand{\gao}{\ga^\phi}

\newcommand{\qo}{q_o}
\newcommand{\qop}{q_\op}
\newcommand{\qto}{q_{to}}
\newcommand{\qio}{q_{io}}
\newcommand{\qtop}{q_{to'}}
\newcommand{\Qo}{Q_o}
\newcommand{\Ao}{A_o}
\newcommand{\Qto}{Q_{to}}
\newcommand{\Ato}{A_{to}}
\newcommand{\Qnull}{Q_{\phi}}
\newcommand{\Anull}{A_{\phi}}
\newcommand{\Qtnull}{Q_{t\phi}}
\newcommand{\Atnull}{A_{t\phi}}

\newcommand{\ro}{r_o}
\newcommand{\rop}{r_{o'}}

\newcommand{\tA}{\widetilde{A}}
\newcommand{\tQ}{\widetilde{Q}}
\newcommand{\tAt}{\widetilde{A}_t}
\newcommand{\tQt}{\widetilde{Q}_t}
\newcommand{\tAi}{\widetilde{A}_i}

\newcommand{\tx}{\widetilde{x}}
\newcommand{\txo}{\tx_o}
\newcommand{\dtxo}{\dt\txo}

\newcommand{\mO}{\mathcal{O}}

\newcommand{\bl}{{\boldsymbol{\la}}}
\newcommand{\bn}{{\boldsymbol{n}}}
\newcommand{\bN}{{\boldsymbol{N}}}
\newcommand{\blo}{\bl_o}
\newcommand{\bno}{{\bn_o}}
\newcommand{\bNo}{{\bN_o}}

\newcommand{\uow}{u_{ow}}
\newcommand{\uww}{u_{ww}}
\newcommand{\uoo}{u_{oo}}

\newcommand{\uwv}{u_{wv}}
\newcommand{\uov}{u_{ov}}
\newcommand{\bu}{\overline{u}}
\newcommand{\buww}{\bu_{ww}}
\newcommand{\buow}{\bu_{ow}}
\newcommand{\buoo}{\bu_{oo}}
\newcommand{\buwv}{\bu_{wv}}
\newcommand{\buov}{\bu_{ov}}
\newcommand{\Dtbuoo}{\Dt \bu_b}%{\Dt\buoo}
\newcommand{\Dtbui}{\Dt\bu_i}
\newcommand{\Dtbub}{\Dt\bu_b}

\newcommand{\bupol}{\bu_{\rm pol}}

\newcommand{\bub}{\overline{u}_b}

\newcommand{\zb}{z_b}
\newcommand{\zi}{z_i}

\newcommand{\ur}{u_r}
\newcommand{\ut}{u_t}
\newcommand{\utr}{u_{tr}}

\newcommand{\bR}{{\mathbf{R}}}

\newcommand{\Unull}{U_\phi}

\newcommand{\cS}{{\mathbb{S}}}

\newcommand{\noo}{n_{oo}}

\newcommand{\QD}{Q_{\rm D}}

\newcommand{\dA}{\dt A}

\newcommand{\dAcomb}{\dA_{\rm comb}}
\newcommand{\dAint}{\dA_{\rm int}}

\newcommand{\ba}{\overline{a}}

\newcommand{\dtbathry}{\dt\ba_{\rm thry}}
\newcommand{\dtbaex}{\dt\ba_{\rm num}}

\newcommand{\bveoo}{\overline{\ve}_{oo}}

\newcommand{\bmu}{\overline{\mu}}
\newcommand{\bmuost}{\bmu_o^*}
\newcommand{\bmuwo}{\bmu_w^\phi}
\newcommand{\bmuw}{\bmu_w}
\newcommand{\bmuo}{\bmu_o}

\newcommand{\bga}{\overline{\ga}}
\newcommand{\bdtga}{\dt\bga}

\newcommand{\Anum}{A_{\rm ex}}	%{A_{\rm num}}
\newcommand{\anum}{a_{\rm ex}}
\newcommand{\ath}{a_{\rm thry}}
\newcommand{\aint}{a_{\rm int}}
\newcommand{\baint}{\ba_{\rm int}}

\newcommand{\dAnum}{\dt\Anum}
\newcommand{\rst}{{\text{RST}}}
\newcommand{\Arst}{A_\rst}

\newcommand{\xot}{x_{ot}}

\graphicspath{{./Figs/},{./}}

\newif\ifjusttheory
\justtheoryfalse

\newif\ifshowfigures
\showfigurestrue

\begin{document}

\title{%A simple model relating solute interactions to interfacial properties}
	A simple theory for interfacial properties of dilute solutions}
\thanks{The following article has been submitted to J.~Chem.~Phys.}
\author{Varun Mandalaparthy}
\author{W.~G. Noid}
\email{wnoid@chem.psu.edu}

\affiliation{Department of Chemistry,  Penn State University,  University Park, Pennsylvania 16802 USA}
\date{\today}

\begin{abstract}
Recent studies suggest that cosolute mixtures may exert significant non-additive effects upon protein stability. 
The corresponding liquid-vapor interfaces may provide useful insight into these non-additive effects. 
%The corresponding liquid-vapor interfaces may provide useful insight into the interactions between solutes and their role in these non-additive effects. 
Accordingly, in this work we relate the interfacial properties of dilute multicomponent solutions to the interactions between solutes. 
We first derive a simple model for the surface excess of solutes in terms of thermodynamic observables. We then develop a lattice-based statistical mechanical perturbation theory to derive these observables from microscopic interactions. Rather than adopting a random mixing approximation, this dilute solution theory (DST) exactly treats solute-solute interactions to lowest order in perturbation theory. Although it cannot treat concentrated solutions, Monte Carlo (MC) simulations demonstrate that DST describes dilute solutions with much greater accuracy than regular solution theory. Importantly, DST emphasizes an important distinction between the ``intrinsic’’ and ``effective’’ preferences of solutes for interfaces. DST predicts that three classes of solutes can be distinguished by their intrinsic preference for interfaces. While the surface preference of strong depletants are relatively insensitive to interactions, the surface preference of strong surfactants can be modulated by interactions at the interface. Moreover, DST predicts that the surface preference of weak depletants and weak surfactants can be qualitatively inverted by interactions in the bulk. We also demonstrate that DST can be extended to treat surface polarization effects and to model experimental data. MC simulations validate the accuracy of DST predictions for lattice systems that correspond to molar concentrations.	
%\\ \\ \vspace{10pt}
%The following article has been submitted to J.~Chem.~Phys.
\end{abstract}

\maketitle

%The following article has been submitted to J.~Chem.~Phys.

%\begin{comment}
\section{Introduction}			
Cosolutes dramatically impact the conformational equilibria of proteins and other macromolecules.\cite{yancey1979, yancey1982, wu2006,rydeen2018} 
For instance, urea and other molecules that interact preferentially with unfolded conformations denature proteins.\cite{nozaki1963, felitsky2004, stumpe2007, guinn2011,canchi2013} 
Conversely, betaine and other molecules that are preferentially depleted from unfolded conformations stabilize proteins.\cite{mondal2013, lin1994, wang1997, larini2013}
%Urea and other denaturants interact preferentially with unfolded protein conformations,\cite{nozaki1963, felitsky2004, stumpe2007, guinn2011,canchi2013} while stabilizers, such as trimethylamine oxide (TMAO) and betaine, tend to be preferentially depleted from folded conformations.\cite{mondal2013, lin1994, wang1997, larini2013} 
However, although proteins function in complex multicomponent solutions, relatively few studies have probed the influence of cosolute mixtures upon protein equilibria.\cite{lin1994,mello2003,holthauzen2006} 
Important early studies concluded that mixtures of denaturants and stabilizers impact proteins in an additive manner.\cite{lin1994,mello2003,holthauzen2006,auton2004,auton2007}  
More recent studies, though, suggest that cosolute mixtures can exert significant non-additive effects upon protein stability.\cite{ganguly2020,narang2017,hunger2015,bruce2019}

%\textbf{Will: The sentence above about osmolytes interacting preferentially with the folded state is \textit{seemingly} at odds with the general osmophobic mechanism of stabilization. Although you are correct in terms of the Wyman-Tanford explanation of stabilization, people may not agree with your statement as written.}

The liquid-vapor interface of aqueous solutions may possibly provide useful insight into the fundamental mechanisms by which cosolute mixtures influence protein stability. 
Clearly, there exist important differences between the interfaces formed by an aqueous phase with a vapor phase and with a protein surface.\cite{southall_view_2002,auton2006} 
Nevertheless, water molecules lose strong hydrogen-bonding interactions at both the liquid-vapor interface and also at hydrophobic protein surfaces.\cite{chandler2005, kauzmann1959} 
Accordingly, many studies have employed the liquid-vapor interface as a relatively simple surrogate for an idealized hydrophobe in order to investigate solvation contributions to protein stability.\cite{arakawa1983, arakawa1985,kita1994,lin1996,lee1981, kaushik1998,liao2017} 
%to the thermodynamic driving forces that govern protein conformational equilibria.\cite{arakawa1983, arakawa1985,kita1994,lin1996,lee1981, kaushik1998,liao2017} 
Moreover, the Gibbs-Duhem equation provides a well-known and rigorous analogy between the influence of cosolutes upon the liquid-vapor interface and their influence upon the conformational equilibria of dilute proteins.\cite{parsegian_osmotic_2000,pegram2009} 
In this analogy, the surface tension corresponds to the Gibbs potential for the conformational transition, while the interfacial excess corresponds to the preferential interaction coefficient.\cite{kita1994,lee1981} 
%Therefore, the interfacial properties of osmolyte mixtures may possibly provide insight into osmolyte interactions and their influence upon macromolecular stability.

More generally, aqueous interfaces profoundly impact many processes of fundamental, industrial, and technological interest.\cite{freedman2017,zdziennicka2020,baccile2021}
Consequently, although they have been studied for centuries,\cite{gibbs1874,rowlinson2002, ono1960} the interfacial properties of aqueous solutions remain the subject of intense research today.\cite{benjamin2015,malila2018,lbadaoui-darvas2022}
%the interfaces of aqueous solutions have been intensely studied for centuries\cite{gibbs1874,rowlinson2002, ono1960} and remain the subject of very active research.\cite{benjamin2015,malila2018,lbadaoui-darvas2022} 
%there is an extremely large body of research that extends over centuries\cite{gibbs1874,rowlinson2002, ono1960}
%and continues today.\cite{benjamin2015,malila2018,lbadaoui-darvas2022} 
Many studies have employed Gibbsian thermodynamics to investigate the fundamental thermodynamic relations governing the surface tension and surface excess of multicomponent solutions.\cite{butler1932, santos2014, kaptay2018, elliott2020}
More pragmatically, many studies have invoked empirical mixing rules or approximate equations of state to relate the interfacial properties of solutions to the properties of pure liquids.\cite{sprow1966, shereshefsky1967,lamperski1991, nath1999, chunxi2000, fainerman2001, tahery2005, brocos2007,janczuk2021}
These approaches often describe liquid-vapor interfaces with binding isotherms in analogy to adsorption on solid interfaces.\cite{connors1989, fainerman1998, li2001, varela2009, wang2011, topping2010,shardt2017}
%interfacial region in analogy to the adsorption of 
%In many cases, these theories invoke Langmuir or other binding isotherms to describe surface adsorption.\cite{connors1989, fainerman1998, li2001, varela2009, wang2011, topping2010,shardt2017}

Other studies have employed statistical mechanics and molecular simulations to relate the interfacial properties of aqueous solutions to molecular interactions.\cite{prigogine1956,stephan_enrichment_2020}
For instance, many studies employ mechanical expressions\cite{kirkwood_statistical_1949, irving_statistical_1950, carey_gradient_1978} to determine the interfacial tension from molecular dynamics simulations.\cite{ghoufi_computer_2016} 
%For instance, early work by Prigogine and coworkers combined a cell-based model of liquids with the law of corresponding state to model the interfacial properties of dilute solutions.\cite{prigogine1956} 
%More recently, mechanical expressions\cite{kirkwood_statistical_1949,irving_statistical_1950,carey_gradient_1978} are often employed to estimate the interfacial tension from molecular dynamics (MD) simulations.\cite{ghoufi_computer_2016}
Many other studies employ local Cahn-Hilliard free energy functionals\cite{cahn1958} to model the interfacial density profile and surface tension.\cite{miqueu_modeling_2005,nino-amezquita_measurement_2010,mairhofer_modeling_2017}
%Following the work of Cahn and Hilliard,\cite{cahn1958} many studies have employed a local free energy with square gradient contributions to model the density profile and surface tension of liquid interfaces.\cite{miqueu_modeling_2005,nino-amezquita_measurement_2010,mairhofer_modeling_2017}
Classical density functional theories\cite{evans_nature_1979,wu_density_2006} are also widely adopted to model the interfaces of aqueous solutions.\cite{llovell_classical_2010,rehner_surfactant_2021}
More recently, Dutcher and coworkers have extensively employed lattice models and random mixing approximations to study the interfacial properties of binary solutions.\cite{dutcher2010,wexler2013,boyer2016,boyer2017,liu2021}

Statistical mechanical perturbation theory provides another approach for investigating interfacial properties. 
%In the present work, we develop a simple statistical mechanical perturbation theory to investigate the interfacial properties of dilute solutions.
% influence of dilute solutes and, in particular, their interactions upon interfacial properties.
This approach was pioneered by Guggenheim, who employed a grand canonical formalism to model interfaces in analogy to adsorption at a solid surface.\cite{guggenheim1945}
However, Guggenheim's treatment of the interface as an independent, autonomous layer was not consistent with the Gibbs adsorption equation and did not properly describe the surface excess.\cite{defay1950,murakami1951,defay1966}
Subsequently, Ono and Kondo developed a grand formalism that determined the influence of solutes upon the interfacial properties of an inhomogeneous system\cite{ono1960} in analogy to the McMillan-Mayer theory for the osmotic pressure.\cite{mcmillan1945,hill1986} 
%[Ono and Kondo 1960] 
%This formalism is analogous to the McMillan-Mayer theory for determining the osmotic pressure due to the addition of a dilute solute.\cite{mcmillan1945,hill1986} 
%Ono and Kondo,\cite{ono1960}  as well as subsequent studies that employed similar formalisms,\cite{prigogine1980,kalies2004}
%then employed a random mixing approximation to treat intermolecular interactions, in analogy to regular solution theory.

In the present work, we employ perturbation theory to investigate the interfacial properties of dilute solutions.
We develop a lattice-based framework that is conceptually very similar to, although perhaps simpler than, the important early work of Ono and Kondo.\cite{ono1960}
In particular, rather than emphasizing general formalism, we focus on the interactions between solutes and their impact upon observable thermodynamic parameters.
Moreover, in contrast to prior perturbative approaches,\cite{guggenheim1945, ono1960, prigogine1980, kalies2004} we do not employ the random mixing approximation that is adopted by regular solution theories for treating the interactions between solutes.\cite{hill1986}
Rather we treat these interactions exactly to their lowest order in perturbation theory. 
Consequently, the resulting theory describes interactions in dilute solutions much more accurately than regular solution theory. 
Moreover, the simplicity of the analytic, lattice-based perturbation theory provides transparent insight into the influence of dilute solutes and, in particular, their interactions upon interfacial properties.
Furthermore, we demonstrate that interfacial polarization effects can be incorporated into this formalism. 
While the present work adopts a lattice model, we anticipate that the general formalism should readily extend to more realistic off-lattice models.

\begin{comment}
In contrast to these prior studies, we do not adopt a random mixing approximation to describe the interactions between solutes. 
Rather, we exactly treat the interactions between solutes to the lowest order in perturbation theory. 
Consequently, the resulting theory describes dilute solutions much more accurately than regular solution theory. 
Moreover, the simplicity of the analytic perturbation theory provides transparent insight into the influence of dilute solutes and, in particular, their interactions upon interfacial properties.
While the present work adopts a lattice model, we anticipate that the general formalism should readily extend to more realistic off-lattice models.
\end{comment}

The remainder of this manuscript is organized as follows.
The theory section initially reviews thermodynamic properties of interfaces, derives a simple expression for the surface excess in terms of experimental observables, and recollects their relationship to the conformational transitions of macromolecules. 
The theory section then develops a perturbation theory that relates the bulk and interfacial properties of dilute solutions to one- and two-solute partition functions. 
The results and discussion section first validates the accuracy of this dilute solution theory and identifies three classes of solutes that can be distinguished by their intrinsic preference for the interface.
This section next demonstrates that repulsive interactions can convert intrinsic depletants to effective surfactants, while attractive interactions can convert intrinsic surfactants to effective depletants. 
The section then investigates the impact of surface polarization upon interfacial properties and relates the formalism to experimental measurements for osmolytes that are commonly employed to modulate protein stability. 
The conclusions section briefly summarizes the manuscript and outlines promising future directions.
The manuscript closes with a detailed methods section describing the numerical calculations. 
A supplementary materials section provides additional results and details from our numerical calculations.

%%% THIS IS THE END OF THE OLD INTRODUCTION

% curvature of the surface tension, given the intrinsic interfacial preference and the bulk interaction parameter and make predictions about the preferential interaction of these osmolytes at interfaces. Specific applications to the well-studied osmolyte urea, betaine, proline, and trimethylamine N-oxide (TMAO) are presented.

%\begin{itemize}
%\item Relating the surface tension to microscopic interactions.
%\item Microscopic model that we simulate exactly.
%\item Rigorous theory that is exact at low concentration
%\item Not assuming an isolated interface, easy to generalize to off-lattice.
%\item Not assuming a regular solution or any mean field approximation. 
%\item Air-water interface is like the ideal hydrophobic polymer interface.
%\end{itemize}

%In this paper, we derive a thermodynamic theory for interfaces based on the lowest order expansion in concentration for the chemical potentials of the solvent and solute. This theory introduces phenomenological constants that we study in detail via a lattice-based statistical mechanics perspective. We show that the space of binary solutions can be completely characterized in terms of two parameters: the solute interfacial preference and the solute aggregation propensity. We identify several interesting regimes and examine where the theory performs well and where our assumptions start to break down. We also analyze experimentally studied binary solute solutions via our framework.

%\newpage

%\end{comment}

\section{Theory}
\label{Sec-Theory}
We first review basic thermodynamic properties of interfaces.
In particular, we relate the surface excess, $\Ga$, to the bulk interaction parameter, $\ve$, and to the sensitivity, $\ptl \ga/\ptl m$, of the surface tension, $\ga$, to the bulk molality, $m$. 
We then develop a statistical mechanical theory for deriving these thermodynamic parameters based upon microscopic interactions.

\subsection{Thermodynamic definition of the preferential adsorption coefficient}
We first consider a homogeneous solution with a single dominant solvent (i.e., water) and multiple distinct solute species, which we label by $o$.
The solution contains $\nw$ water molecules and $\no$ $o$-solute molecules.
We shall implicitly assume that the temperature and pressure are constant throughout this subsection. 
The relevant free energy is the Gibbs potential, $G = \nw\muw + \sumo \no\muo$, where $\muw$ is the water chemical potential, $\muo$ is the chemical potential for $o$-solutes, and the summation is performed over the various solute species.
The Gibbs-Duhem relationship for the bulk solution is $\dmuw = - \sumo \mo \dmuo$, where $\mo = \no / \nw$ is the dimensionless molality.
%We begin with a general thermodynamic description of a bulk solution at constant temperature and pressure with $n_w$ water molecules and $n_o$ solute molecules of type $o$. The Gibbs free energy, $G = n_w \mu_w + \sum_o n_o \mu_o$ for this system introduces the chemical potentials of the solute, $\mu_o$, and the water, $\mu_w$. At constant temperature and pressure, these chemical potentials are related via the Gibbs-Duhem relationship, $d\mu_w=-\sum_o m_o d\mu_o$. Here, $m_o$ represents the molality of the solute species, defined as $n_o/n_w$. We express these chemical potentials as the lowest order expansions in the concentrations to include interactions:
%\begin{equation}\label{eq:mu_w}
%\mu_w = \mu_w^0 - \kbt \sum_i  m_i -\half[  \ve_{20} m_1^2 -2\ve_{11} m_1 m_2 - \ve_{02} m_2^2 ]+ \mathcal{O}(m^3)
%\end{equation}
We expand the chemical potentials as a perturbation theory in solute molality:
% and systematically truncate at the lowest order that incorporates solute-solute interactions:
\begin{eqnarray}
\label{eq:mu_w}
\muw 
	& = & 
		\muwo 
	- 	\kbt \sumo  \mo 
	-	\half \sum_{o,o'} m_o \ve_{oo'} m_{o'}  + \mO(m^3)	\\
\label{eq:mu_i}
\mu_o 
	& = & 
		\muo^* 
	+ 	\kbt \ln m_o 
	+ 	\sumop \ve_{\oop} \mop +  \mO(m^2)	.
\end{eqnarray}
Here $\muwo$ is the chemical potential of pure water and $\mu_o^*$ is related to the solvation free energy of the solute.
The second terms in Eqs.~\eqref{eq:mu_w} and \eqref{eq:mu_i} correspond to the van't Hoff law for osmotic pressure and the translational entropy of ideal solutes, respectively. 
The third terms account for solute-solute interactions in terms of the fixed energetic parameter, $\ve_{oo'}$. 
%These definitions are consistent with the Flory-Huggins free energy (details in the SI). 
While Eqs.~\eqref{eq:mu_w} and \eqref{eq:mu_i} adopt the conventional form of regular solution theory, they more generally correspond to the lowest order perturbation theory that treats solute-solute interactions.
%a perturbation theory in solute concentration at the lowest order that incorporates solute-solute interactions.

%This truncation of the chemical potentials is justified when the solutions are sufficiently dilute, and we therefore expect our theory to be most accurate at lower solute concentrations.

We next consider an inhomogeneous solution of $\Nw$ water molecules and $\No$ $o$-solutes at the same temperature and pressure.
The Gibbs potential for this inhomogeneous system is $G_t = \Nw \muw + \sumo \No \muo + \ga \sigma$, where $\ga$ is the interfacial tension and $\sigma$ is the surface area.
The Gibbs-Duhem relation for the inhomogeneous system may be expressed $\dga = - \sigma\inv\left\{ \Nw\dmuw + \No\dmuo \right\}$.
After employing the bulk Gibbs-Duhem equation to eliminate $\dmuw$, one obtains the Gibbs adsorption equation:
\begin{equation}
\label{eq:adsorption_equation}
\dga = -\sumo \Gao \dmuo
\end{equation}
where  
\begin{equation}
\Gao = \sigma\inv \left(  \No - \mo \Nw \right)
\end{equation}
is the surface excess of $o$-solutes and $\mo$ is the molality of the bulk solution in chemical equilibrium with the interface.
%
%Here, $\sigma$ is the surface area of the interface while $N_{oi}$ and $N_{wi}$ represent the number of solute and water molecules at the interface respectively.
We next express the surface tension as a power series in molality and again truncate at the lowest order that treats solute-solute interactions:
%\begin{equation} \label{eq:surften_fit}
%\overbar{\gamma} - \overbar{\gamma}_0 = \alpha_1 m_1 + \alpha_2 m_2 + \half \sqbracket{\beta_{20}m_1^2 + 2 \beta_{11}m_1m_2 +  \beta_{02}m_2^2} + \mathcal{O}(m^3)
%\end{equation}

%\begin{equation} \label{eq:surften_fit}
%\overbar{\gamma} - \overbar{\gamma}_0 = \sum_o \alpha_o m_o + \half \sum_{o,o'} m_o \beta_{oo'}m_{o'} + \mathcal{O}(m^3)
%\end{equation}
\begin{equation} 
\label{eq:surften_fit}
%\bt \left( \ga - \ga^\phi \right)  
\bt \ga
	=
		\bt \ga^\phi 
	-	\sumo \ko \mo 
	+ \half \sumoop \mo h_\oop \mop + \mO(m^3)	,
\end{equation}
where $\bt = 1/\kT$ and $\ga^\phi$ is the surface tension of pure water, while $\ko \equiv - \left. \ptl \bt\ga/\ptl \mo \right|_\phi$ and $h_\oop \equiv  \left. \ptl^2 \bt\ga / \ptl \mo \ptl \mop \right|_\phi$ are the first and second derivatives of the surface tension at infinite dilution.

The expressions for $\muo$ in Eq.~\eqref{eq:mu_i} and for $\ga$ in Eq.~\eqref{eq:surften_fit} determine a model for the surface excess.
We define the susceptibility matrix 
\begin{equation}
\label{def-Goop}
G_\oop 
	\equiv 
		\left( 
			\frac{\ptl \left( \bt \muo \right) }{\ptl \mop} 
		\right)_{m_\op^*}
	= 
		\mo\inv \delta_\oop + \bt \ve_\oop + \mO(m)	.
\end{equation} 
We also define 
\begin{equation}
\label{def-bo}
b_o 
	\equiv 
		\left(
			\frac{\ptl \left( \bt \ga \right) }{\ptl \mo} 
		\right)_{m_o^*}
	= 
		- \ko + \sumop h_\oop \mop + \mO(m^2), 
\end{equation}
which quantifies the sensitivity of the surface tension to the bulk molality of $o$-solutes.
The subscript $m_o^*$ indicates that the molality of other solutes is held constant in the partial derivative. 
The Gibbs adsorption equation then implies that
\begin{equation}
b_o = - \sumop G_\oop \Ga_\op	.
\end{equation}
This equation can be inverted to determine the vector of surface excesses:
\begin{equation}
\label{eq-ThermoGa}
\Ga = - G\inv b.
\end{equation}
If we treat $G$ and $b$ to lowest order, one obtains relatively simple and instructive expressions for $\Ga$.
In particular, if the solution contains only a single solute species with molality $m$, then
\begin{equation}
\label{eq-Gao-one}
\Ga = m \left( \frac{k - h m}{1 + \ove m} \right)	
\end{equation}
where $\ove = \ve/\kT$.
If instead the solution contains two solute species, 1 and 2, then the surface excess of solute 1 may be expressed
\begin{equation}
\label{eq-Gao-two}
\Ga_1 = m_1 
		\left[ 
			\frac{k_1 - h_{11} m_1 - (h_{12} - k_1 \ove_{22}+ k_2 \ove_{12}) m_2}
				{1 + \ove_{11} m_1 + \ove_{22} m_2} 
		\right]	
\end{equation}
and a corresponding expression describes $\Ga_2$. 
Thus, the parameter $\ko$ describes the intrinsic tendency of $o$-solutes to accumulate at the interface.
However, Eq.~\eqref{eq-Gao-two} demonstrates that interactions with other solutes can reverse this tendency and, e.g., drive surfactants from the surface.
%These expressions are quite reminiscent of Kirkwood-Buff theory,\cite{ben-naim2006, smith2006} which is derived for open solutions and applies in any concentration regime.
%In contrast, the present framework applies to closed solutions and is only valid for dilute solutions. 

\newcommand{\muM}{\mu_{\rm M}}

It is worth briefly recalling the close connection between the thermodynamics of interfaces and dilute macromolecules.\cite{parsegian_osmotic_2000}
In particular, if $\Dt \muM = \mu_{\rm F} - \mu_{\rm U}$, is the difference in the chemical potentials of the F and U conformations of a dilute protein, then the same reasoning leads to an analogous Gibbs-Duhem equation for the free energy of the conformational transition:
\begin{equation}
\dd\Dt\muM = - \sumo \Dt \Gao \dmuo	,
\end{equation}
where $\Dt\Gao = \Ga_{{\rm F}|o} - \Ga_{{\rm U}|o}$ quantifies the preference of $o$-solutes for the F conformation relative to the U conformation.
Furthermore, the sensitivity of the conformational transition to $o$-solutes, $b_{{\rm M}|o}  = \ptl(\bt\Dt\muM)/\ptl \mo$, can be related in a similar linear system of equations to the preferential adsorption: $\Dt\Ga = - G\inv b_{\rm M}$, which gives a result analogous to Eq.~\eqref{eq-Gao-two} in the case of two solutes.
Consequently, this framework may possibly be useful for understanding the influence of cosolute interactions upon macromolecular conformational transitions.
Specifically, $\Dt\Gao$ can be inferred from experimental measurements of bulk thermodynamics, $G$, and conformational stability, $b_{{\rm M}|o}$,\cite{rosgen_molecular_2007,rosgen_synergy_2015} which may then provide insight into potential non-additive effects of cosolutes upon macromolecules.

\subsection{Microscopic theory}
\label{SubSec-StatMech}
We now develop a statistical mechanical theory for the influence of dilute solutes upon interfaces.
We first adapt Hill's constant pressure solution theory\cite{hill_theorysolns_1957} to derive microscopic expressions for the thermodynamic solute-solute interaction parameters, $\veoop$ and $\ve_{t\oop}$, describing homogeneous and inhomogeneous systems, respectively.
Hill's formalism is analogous to the virial expansion for the pressure of dilute gases and to McMillan-Mayer theory for dilute solutions, but is performed at constant pressure rather than constant volume.\cite{hill1986}
%Hill's constant pressure theory is analogous to the standard virial expansion for the pressure of dilute gases, but the perturbation expansion is performed in terms of the concentration of dilute solutes.
By eliminating the bulk contributions to the free energy of the inhomogeneous system, we then derive microsocopic expressions for the interfacial tension, $\ga$, and surface excess, $\Ga$, as a function of the bulk composition.
In the present work, we adopt a simple lattice model both because it makes microscopic expressions for the thermodynamic parameters particularly transparent and also because it facilitates numerical assessment of the theory.
However, due to the generality of Hill's formalism, it should be possible to extend our results for the lattice model to more realistic off-lattice models.

\begin{comment}
We now extend Hill's constant pressure solution theory to develop a microscopic theory for the influence of dilute solutes upon interfaces. 
In the present work, we develop a simple lattice theory because it makes microscopic expressions for the thermodynamic parameters particularly transparent.
However, the generality of Hill's formalism ensures that our results for the lattice model can be readily adapted to more realistic off-lattice models. 
\end{comment}

\subsubsection{Homogeneous system}
\label{SubSec-StatMechHomogeneousSystem}
We first consider a lattice model for a homogeneous multicomponent mixture of $\nw$ water molecules and $\no$ $o$-solute molecules.
The system is described by $M$ equivalent lattice sites each with a fixed volume, $v_1$, such that the total volume of the lattice is $V = v_1 M$.
Each lattice site is occupied by a single molecule such that $M = \nw + \sumo \no$.
We treat the temperature, $T$, as an implicit constant throughout this work.
We denote the bare pressure and chemical potentials of the lattice model by $\hat{P}$ and $\hat{\mu}$, respectively. 
The Helmholtz potential, $A$, is the analog of the Gibbs potential for the lattice model:
\begin{equation}
A = -\hat{P} V + \hmuw \nw + \sumo \hmuo \no = \muw M + \sumo \muo \no	,
\end{equation}
where $\muw = \hmuw - \hat{P}v_1$ is the free energy cost of adding a lattice site, while $\muo = \hmuo - \hmuw$ is the free energy cost of replacing a water molecule with an $o$-solute:
\begin{equation}
\dd A = \muw \dM + \sumo \muo \dno		. 
\end{equation}
%Consequently, $M$ and $\setno$ define the thermodynamic equilibrium state.
The lattice Gibbs-Duhem equation is 
\begin{equation}
\label{eq:lattice_gibbs_duhem}
\dmuw = - \sumo \xo \dmuo			,
\end{equation}
where $\xo = \no/M$ is the mole fraction of $o$-solutes.

We denote the canonical partition function for the homogeneous mixture by $Q(\setno) \equiv Q(M,\setno) = \exp[-\bt A]$.
We define $\Qnull = Q(\{0\})$ and $\Anull = - \kT \ln \Qnull = M \muwo$ as the canonical partition function and Helmholtz potential in the absence of solute, i.e., when $\nw = M$ and $\no = 0$.
We define $\Qo = \left. Q(\{n_{\op}\})\right|_{n_\op = \delta_\oop}$ and $\Ao = -\kT \ln \Qo$ as the partition function and Helmholtz potential in the case that the system contains a single $o$-solute.
% molecule has been added to the system.
We also define
\begin{equation}
\label{def-qo}
\qo = M\inv \Qo/\Qnull	.
\end{equation}
Because there are $M$ equivalent sites to place the solute, it follows that $\qo$ is independent of $M$ or any other thermodynamic variable (other than $T$).
The quantity $\qo$ can be interpreted as an effective partition function associated with introducing an osmolyte into a single fixed lattice site that is surrounded by solvent.

We next suppose that the system is open such that solutes can replace water molecules.
The relevant free energy is then a natural function of $M$ and the solute chemical potentials:
\begin{eqnarray}
\label{def-tA}
\tA 
	& \equiv & A - \sumo \no \muo = \muw M			\\
\dd\tA 
	& = &
		\muw \dM - \sumo \no \dmuo	.
\end{eqnarray}
The relevant partition function for this (semi-) grand ensemble is
\begin{equation}
\label{def-tQ}
\tQ = \exp[-\bt \tA] 
%	= \sumno \lo^\no Q(\setno)	,
	= \sumno \blo^\bno Q(\setno)			,
\end{equation}
where $\blo^\bno = \prod_o \lo^\no$ is a product of the absolute activities, $\lo = e^{\bt\muo}$, for each solute species $o$ and the sum is evaluated over all possible compositions of the system.
We define the ratio of partition functions, $\Psi\equiv\tQ/\Qnull$, and expand this ratio in the solute activity, $a_o \equiv \qo \lo$:
\begin{eqnarray}
\Psi 
	&\equiv & \frac{\tQ}{\Qnull} 
		=
			e^{-\bt M \dtmuw } 		%\\ \nonumber
%&=&
		= 
1 + M \sum_o a_o  + \half \sum_{o,o'} a_o Z_{oo'} a_{o'} + \mO(a^3)	,
\end{eqnarray}
%
%\textbf{DEFINE $\delta\wt{\mu}_w$.}
where $\dtmuw = \muw - \muwo$ is the change in the water chemical potential due to the addition of solutes.
We have also defined
\begin{equation}
Z_{oo'} = \frac{C_{oo'}}{\qo \qop}  \frac{Q^{(2)}_{oo'}}{ \Qnull}	.
\end{equation}
where $Q^{(2)}_{oo'}$ is the canonical partition function $Q(\setno)$ in the case that the solution contains exactly 2 solute molecules with one being of type $o$ and the second being of type $o'$, while $C_{oo'}$ is a combinatorial factor such that $C_{oo} = 2 $ and $C_{oo'} = 1$ if $o\neq o'$.
Taking the (natural) logarithm of $\Psi$ and expanding in $\ao$, we arrive at:
\begin{equation}
\label{eq:lnPsi}
-\bt \dtmuw
	= \frac{1}{M}\ln\Psi 
	=
		\sumo \ao
	+ 
		\half \sum_{o,o'} a_o \varphi_{oo'}  a_{o'} 
	+
		 \mO(a^3)	,
\end{equation} 
where
\begin{equation}
\label{def-varphioo}
\varphi_{oo'} = M\inv(Z_{oo'} - M^2)	.
\end{equation} 
The lattice Gibbs-Duhem equation (Eq.~\eqref{eq:lattice_gibbs_duhem}) implies that
\begin{equation}
\xo 
	= 
		\ao \left( \frac{\ptl (-\bt \dtmuw)}{\ptl \ao} \right)_{a_o^*}
	= 
		\ao 
			\left\{ 
				1 + \sum_{o'} \varphi_{oo'} a_{o'} + \mO(a^2)
			\right\}	.
\end{equation}	
This can be inverted to obtain:
\begin{equation}
\label{eq:ao}
\ao = \xo \left\{ 1 - \sum_{o'} \varphi_{oo'} x_{o'} + \mO(x^2) \right\}	,
\end{equation}
where the quantity in brackets defines the activity coefficient.
%This then allows us to introduce the activity coefficient via $a_o = f_o x_o$:
%%
%\begin{equation}\label{eq:activity_coefficient}
%f_o = 1 - \varphi_{oo}x_o - \varphi_{ij}x_{j} + \mathcal{O}(x^2)
%\end{equation}
This expansion then leads to the lattice analogs of  Eqs.~\eqref{eq:mu_w} and \eqref{eq:mu_i}:
\begin{eqnarray}
\muw 
	& = & 
		\muwo - \kT \sumo \xo - \half \sum_{o,o'} x_o \ve_{oo'} x_{o'} + \mO(x^3)	\\
\muo 
	& = &
	 	\muost + \kT \ln \xo + \sumop \veoop \xop + \mO(x^2)	,
\end{eqnarray}
where 
\begin{eqnarray}	
\label{def-muost}
\mu_o^* & = & - \kT \ln q_o		\\
\label{def-veoo}
\ve_{oo'} & = & - \kT \varphi_{oo'} .	%= -\kT M\inv \left( Z_{oo'} - M^2 \right)	.
\end{eqnarray}

\begin{comment}	%Varun's original
\begin{equation}\label{eq:mu_w_lattice}
\delta \wt{\mu}_w = \wt{\mu}_w - \wt{\mu}_{w\phi} = -\kbt \sum_o x_o 
-\half \sum_{o,o'}  x_o \ve_{oo'} x_{o'}  + \mathcal{O}(x^3)
\end{equation}
%
\begin{eqnarray}\label{eq:mu_o_lattice}
\wt{\mu}_o 
&=& \wt{\mu}_o^* + \kbt \ln a_o \\ \nonumber
&=& \wt{\mu}_o^* + \kbt \ln x_o - \sum_{o'}\ve_{oo'} x_{o'} + \mathcal{O}(x^2)
\end{eqnarray}
%
We therefore obtain an expression for the interaction parameter, $\ve_{oo'}$:
%
{\color{blue}
\begin{equation}\label{eq:bulk_epsilon}
\overbar{\ve}_{oo'} = \beta \ve_{oo'}
=
-\varphi_{oo'}
=
M^{-1}(M^2 - Z_{bN_1N_2} ) 
\end{equation}
}
%
Our theory, therefore, provides a way to obtain the bulk interaction coefficient if the two-particle bulk partition function can be calculated. This partition function explicitly accounts for the energies of the different microstates and is not a mean field quantity (Fig. \ref{fig:heatmaps_bulk}).
\end{comment}

%\newpage
\subsubsection{Inhomogeneous system}
\label{SubSec-StatMechInhomogeneousSystem}
We next consider a lattice model for an inhomogeneous system consisting of both bulk and interfacial regions.
The lattice model consists of $M$ lattice sites that are occupied by $\Nw$ water molecules and $\No$ $o$-solutes.
We assume that $M_b$ of these sites are equivalent bulk sites, while $M_i = M - M_b$ are equivalent interfacial sites. 
We define $\sigma_1$ as the surface area of a single interfacial site, such that the surface area of the system is $\sigma = M_i \sigma_1$.
% such that the total surface area is $\sigma = M_i \sigma_1$.
We denote the total Helmholtz potential for this inhomogeneous system by $A_t \equiv A_t(\setNo,M,M_i) \equiv A_t(\setNo):$ 
\begin{eqnarray}
A_t & = & 
		-\hat{P} V 	+ \hmuw \Nw + \sumo \hmuo \No + \hgamma \sigma
		= \muw M + \sumo \muo \No + \gamma M_i,		\\
\dd A_t 
	& = & 
		\muw \dM + \sumo \muo \dNo + \gamma \dM_i		,
\end{eqnarray}
where $\hgamma$ is the bare surface tension of the lattice and $\gamma = \sigma_1 \hgamma$ is the scaled surface tension quantifying the free energy cost of converting a bulk site to an interfacial site.
In particular, we define $\gamma^\phi$ as  the surface tension for pure solvent and $\Atnull = \left. A_t\right|_{\No = 0} = M \muwo + M_i \gamma^\phi$ as the Helmholtz potential for the inhomogeneous pure solvent system.
We define $\Ato = \left. A_t \right|_{\Nop = \delta_{o,o'}}$ as the Helmholtz potential when a single $o$-solute has been introduced into the inhomogeneous system.
The lattice Gibbs-Duhem equation for the inhomogeneous system is
\begin{equation}
\label{eq:inhomogeneous_gibbs_duhem}
\dga = M_i\inv \left( -M \dmuw - \sumo \No \dmuo \right) 	.
\end{equation}

We denote $Q_t(\setNo) \equiv \exp[-\bt A_t(\setNo)]$ as the canonical partition function for a closed inhomogeneous system with the specified composition. 
%In particular, $A_t = - \kT \ln Q_t(\setNo)$ is the Helmholtz potential for an arbitrary mixture.
%We define $\Qtnull \equiv \exp[-\bt \Atnull]$ and $\Qto \equiv \exp[-\bt \Ato]$
We denote $\Qtnull \equiv \exp[-\bt \Atnull]$ as the canonical partition function when the inhomogeneous system contains only solvent.
%We define $\Atnull = M \muwo + M_i \gamma^\phi = - \kT \ln \Qtnull$ relates the Helmholtz potential and canonical partition function in the absence of solutes, while $\gamma^\phi$ denotes the corresponding solvent surface tension.
We denote $\Qto \equiv \exp[-\bt \Ato]$ as the canonical partition function when the system contains a single $o$-solute.
%We define $\Ato = - \kT \ln \Qto$ as corresponding quantities in the case that the system contains a single solute of type $o$.
Recalling that $\qo$ is an effective partition function for solvating a single $o$-solute in a fixed bulk site at infinite dilution, we define $\qio$ as the corresponding effective partition function for solvating a single $o$-solute in a fixed interfacial site at infinite dilution.
Then we define
\begin{equation}
\label{def-qto}
\qto 
	\equiv 
		\frac{1}{M} \frac{\Qto}{\Qtnull} 
%	\equiv \qo + \eti \qo \ko	,
	= \qo + \eti \left( \qio - \qo \right),	
\end{equation}
where $\eti = M_i / M$ is the fraction of interfacial sites. 
Note that $\qto$ depends only upon $\eti$ (and $T$) and is independent of $M$, as well as composition and chemical potentials.
\begin{comment}
Since $\qo$ is an effective partition function for solvating a single $o$ solute in a fixed bulk site, we interpret $\qo\ko$ as an effective partition function for solvating a single $o$ solute in a fixed interfacial site.
Consequently, $\ko$ is a (temperature-dependent) constant that gives the ratio between the effective single site partition functions for interfacial and bulk lattice sites at infinite dilution.
We shall see that $\ko$ quantifies the intrinsic preference of $o$ solutes for the interface.
\end{comment}

We next consider that the inhomogeneous system is open.
The relevant free energy is 
\begin{eqnarray}
\tAt 		
	& \equiv & 
		A_t - \sumo \No \muo = \muw M + \gamma M_i	\\
\dd \tAt
	& = & 
		\muw \dM - \sumo \No \dmuo + \gamma \dM_i	.
\end{eqnarray}
The corresponding partition function is 
\begin{equation}
\label{def-tQt}
\tQt = \exp[-\bt\tAt] 
%	= \sumNo \lo^\No Q_t(\setNo)
	= \sumNo \blo^\bNo Q_t(\setNo)
\end{equation} 
where $\blo^\bNo = \prod_o \lo^\No$.
We define the ratio of partition functions for open inhomogeneous systems:
\begin{eqnarray}
\label{eq:def-Psit}
\Psi_t 
	& \equiv & \frac{\tQt}{\Qtnull} 
	= 
		\exp\left[	-\bt \left( M \dtmuw + M_i \dt\gamma \right) \right] %\\ \nonumber
%&=&
	= 
		1 + M \sumo \ato  + \half \sumoop \ato Z_{too'} \atop + \mO(a_t^3)	.
\end{eqnarray}
Here $\dt\gamma = \gamma - \gamma^\phi$ quantifies the influence of solutes upon the surface tension, while $\ato = \qto \lo$ defines a solute activity for the inhomogeneous system. 
The coefficient 
$
Z_{too'} = \frac{C_{oo'}}{\qto \qtop}  \frac{Q^{(2)}_{too'}}{ \Qtnull}
$ 
is defined by the canonical partition function, $Q^{(2)}_{too'}$ , for exactly 2 solute molecules in the inhomogeneous system.
Expanding $M\inv \ln \Psi_t$ from Eq.~\eqref{eq:def-Psit}, we obtain:
%Taking the logarithm of Eq.~\eqref{eq:def-Psit} and expanding, we obtain:
\begin{equation}
\label{eq:lnPsit-expansion}
-\bt \left( \dtmuw + \eti \dt\gamma \right)
	=
		M\inv \ln \Psi_t 
	=
		\sumo \ato
	+ 
		\half \sumoop \ato \varphi_{t oo'}  \atop 
	+
		 \mO(a_t^3)				,
\end{equation} 
where
\begin{equation}
\label{eq:varphit}
\varphi_{t \oop} \equiv M\inv \left( Z_{t oo'} - M^2 \right) 	\equiv - \bt \ve_{t\oop}	,
\end{equation}		
and we have defined an energetic parameter $\ve_{t\oop} = - \kT \varphi_{t\oop}$ that describes the influence of solutes upon the free energy of the inhomogeneous system.
At this point, we could follow Subsection~\ref{SubSec-StatMechHomogeneousSystem} and relate the activity $\ato$ to the mole fraction of solute in the inhomogeneous system $x_{to} = \No/M$.
However, because we are interested in relating interfacial properties to the composition, $\xo$, of the coexisting bulk region, we continue along a different track.

\begin{comment}		%% This is Varun's original
Here, $Q_{t\phi} = Q_t(\{0\})$ is the partition function for the closed total system with no solutes i.e., pure water. $a_{to} = q_{to}\wt{\lambda}_o $ is the activity of the solute in the total system with $\wt{\lambda}_o = e^{\beta\wt{\mu}_o}$. 
We define the  coefficient $Z_{tN_1N_2} = N_1!N_2! Q_{tN_1N_2}/(q_{t1}^{N_1}q_{t2}^{N_2} Q_{t\phi})$. We have further defined a parameter $r_o$ that relates the total system quantities to the bulk system:
%
\begin{equation}\label{eq:r_o}
r_o = r_o (T, M_i/M) = \frac{q_{to}}{q_o}
\end{equation}
%
where $q_{to}= q_{to}(T, M_i/M) = M^{-1}Q_{to}/Q_{t\phi} = r_o q_o(T)$ is a ratio of partition functions for the closed system. 
This in turn relates the solute activity in the total system to the bulk system via $a_{to} =  r_o a_o$.
%
Finally, we arrive at an expression for the total system free energy difference:
%
{\color{blue}
\begin{eqnarray}
\frac{-\beta \delta\wt{A}_t}{M}
&=&
-\beta \sqbracket{\delta\wt{\mu}_w + (M_i/M)\delta\wt{\gamma}} \\ \nonumber
&=& \frac{1}{M} \ln \Psi_t 
= \sum_o a_{to} 
-\half \sum_{o,o'} a_{to} \ve_{too'}  a_{to'} + \mathcal{O}(a_t^3)
\end{eqnarray}
}%
where $\ove_{too'}$ is an interaction parameter for the total system, analogous to the bulk treatment (Eq. \eqref{eq:bulk_epsilon}):
{\color{blue}
\begin{equation}
\ove_{too'} = \beta \ve_{too'}
=
\frac{1}{M}(M^2 - Z_{tN_1N_2} ) 
\end{equation}
%
}
\end{comment}

\subsubsection{Interface}
\label{SubSec-StatMechInterface}
We now relate the open inhomogeneous system to an open homogeneous system with the same number of sites, $M$, and the same solute chemical potentials, $\{\muo\}$.
According to Eq.~\eqref{eq:lattice_gibbs_duhem}, the water chemical potential is also the same in both systems.
We define the interfacial free energy of the open inhomogeneous system, $\tAi$, by the difference between the free energies of the open inhomogeneous system, $\tAt$, and the open bulk homogeneous system, $\tA$:
%We define the interfacial free energy of the open inhomogeneous system as the difference between the free energy of the open inhomogeneous system, $\tAt$, and the free energy, $\tA$, of an open bulk system with the same number of sites, $M$, and the same chemical potentials, $\muw$ and $\muo$:
\begin{eqnarray}
\label{eq:def-tAi}
\tAi 		
	& \equiv & 
		\tAt - \tA = \ga M_i	\\
\label{eq:dtAi}
\dd \tAi 	
	& = & 
		- M_i \sumo \Gao \dmuo + \ga \dM_i		,
\end{eqnarray}
%The interfacial free energy, $\tAi$, is the difference between the free energy of the open inhomogeneous system, $\tAt$, and the free energy, $\tA$, of a homogeneous bulk system with the same number of sites, $M$, and the same chemical potentials, $\muw$ and $\muo$.
%In Eq.~\eqref{eq:dtAi}, 
where we have defined the surface excess of $o$-solutes:
\begin{equation}
\label{eq:defGao}
\Gao \equiv M_i\inv \left( \No - \xo M \right) 	,
\end{equation}
and $\xo = \xo(\setmuo)$ specifies the composition of the homogeneous system.
%which is the surface excess of solute $o$ due to the presence of the interface.
The corresponding Gibbs-Duhem equation for the surface free energy is the lattice Gibbs adsorption equation:
\begin{equation}
\label{eq:GibbsAbsorptEqn}
\dga = - \sumo \Gao \dmuo	.
\end{equation}
We denote the interfacial free energy of pure solvent by $\tA_{i\phi} = A_{i\phi} = M_i \ga^\phi$, such that $\dt\tAi = \tAi - \tA_{i\phi} = M_i \dt \ga $ quantifies the influence of solutes upon the interfacial free energy. 
Equations~\eqref{eq:lnPsi} and \eqref{eq:lnPsit-expansion} then imply 
\begin{equation}
\label{eq-btdt_tAi}
-\bt\dt\tAi = - \bt M_i \dt\ga = \ln \left( \Psi_t / \Psi \right) 	.
\end{equation}

Before proceeding further, we return to Eq.~\eqref{def-qto}
\begin{equation}
\nonumber
\qto 
	\equiv 
		\frac{1}{M} \frac{\Qto}{\Qtnull} 
%	\equiv \qo + \eti \qo \ko	,
	= \qo + \eti \left( \qio - \qo \right),	
\end{equation}	
where $\qo$ and $\qio$ are effective partition functions for an $o$-solute in fixed bulk and interfacial lattice sites, respectively, at infinite dilution and $\eti = M_i/M$ is the surface/volume ratio.
We now define 
\begin{equation}
\label{def-ko}
\ko = \frac{\qio}{\qo} - 1	,
\end{equation}
as the intrinsic preference of $o$-solutes for the interface.
Importantly, $\ko$ is a constant that is independent of all thermodynamic parameters (other than $T$).
It is interesting that Eq.~\eqref{def-ko} implies $\ko \ge -1$.
% and quantifies the intrinsic preference of $o$ solutes for the interface.
Because they are defined to have the same chemical potentials, we can relate the activities of the homogeneous ($\ao$) and inhomogeneous systems ($\ato$):
\begin{equation}
\label{eq:r_k}
\ro \equiv \qto/\qo = \ato/\ao = 1 + \eti \ko, 
\end{equation}
which depends upon both $\ko$ and the surface/volume ratio $\eti$.	% = M_i/M$.

Having defined $\ko$ and $\ro$, we now employ the expansions of Eqs.~\eqref{eq:lnPsi} and \eqref{eq:lnPsit-expansion} in Eq.~\eqref{eq-btdt_tAi} to obtain
\begin{equation}
-\bt \dt\ga = \sumo \ko \ao + \half \sumoop \ao \varphi_{i oo'} \aop + \mO(\ao^3)		,
\end{equation}
where the interfacial interaction parameter is
\begin{equation}
\varphi_{i oo'} = \eti\inv \left( \ro \varphi_{t oo'} \rop - \varphi_{oo'} \right)	.
\end{equation}
%and we have defined
%\begin{equation}
%\ro \equiv \ato/\ao = \qto/\qo = 1 + \eti \ko.
%\end{equation}
Moreover, Eq.~\eqref{eq:GibbsAbsorptEqn} implies 
\begin{equation}
\Gao 
	= 
		\ao \left( \frac{\ptl \left( -\bt\dt\ga \right) }{\ptl \ao}\right)_{a_o^*}	
	=
		\ao \left( \ko + \sumop \varphi_{i oo'} \aop + \mO(\ao^3) \right) 	.
\end{equation}
Finally, we employ Eq.~\eqref{eq:ao} to express $\dt\ga$ and $\Gao$ in terms of the bulk composition, $\xo$:
\begin{eqnarray}
								\label{eq-btdtga-xo}
-\bt \dt\ga 
	& = & 
		\sumo \ko \xo
	- 
		\half \sumoop \xo h_{oo'} \xop
	+ 
		\mO(x^3)			\\
								\label{eq-Gao-xo}
\Gao 
	& = & 
		\xo 
			\left\{ 
				\ko 
			- 	\sumop \left( h_\oop + \kop \bt \ve_\oop \right) \xop 
			+ 	\mO(x^2) 
			\right\} 
\end{eqnarray}
where 
\begin{equation}
\label{def-hoop}
h_\oop 
	= 
		\bt
		\left\{
				\eti\inv \ro \left( \ve_{t\oop} - \ve_\oop \right) \rop 
			+ 	\eti \ko \kop \ve_\oop
		\right\}	.
\end{equation}
%determines the curvature of $\dt\ga$.
Note that $\ko$ and $\ve_\oop$ depend only upon $T$, while $\ro$ and $\ve_{t\oop}$ depend upon both $T$ and also the surface/volume ratio, $\eti = M_i/M$.
For sufficiently large systems, one expects that $\eti\to0$ and $\ve_{t\oop} = \ve_\oop + \dt \ve_{i\oop} \eti + \mO(\eti^2)$, where $\dt \ve_{i\oop}$ is independent of $\eti$.
Consequently, $h_\oop$ should have a well-defined thermodynamic limit, $h_\oop \to \bt \dt \ve_{i\oop}$.

\begin{comment}
The parameter $\ko$ determines the slope of both $\ga$ and $\Gao$ for infinitely dilute solutions.
Infinitely dilute $o$-solutes with $\ko > 0$ accumulate at the interface and decrease the surface tension.
Conversely, infinitely dilute $o$-solutes with $\ko < 0$ are depleted from the interface and increase the surface tension.
Equation~\eqref{def-hoop} relates the curvature, $h_\oop = \ptl^2 \bt\ga/\ptl\xo\ptl\xop$, of the surface tension to microscopic partition functions that reflect solute-solute interactions at infinite dilution. 
Equation~\eqref{eq-Gao-xo} explicitly demonstrates that the presence of $\op$-solutes can modulate the surface excess, $\Gao$, of $o$-solutes.
This effect is related to the curvature of $\ga$, but is also modulated by the intrinsic surface preference, $\kop$, of $\op$-solutes and by the effective interaction, $\ve_\oop$, between $o$- and $\op$- solutes.
Note that Eq.~\eqref{eq-Gao-xo} gives $\Gao$ as a Taylor series expansion to second order in solute concentration.
In comparison, the rational fraction expression for $\Gao$ in, e.g., Eq.~\eqref{eq-Gao-two} is expressed in terms of the same parameters (i.e., $\ve$, $k$, and $h$), but remains valid over a considerably larger composition regime. 
Consequently, we will adopt the rational fraction expression for $\Gao$ in the following.
\end{comment}

This completes our microscopic derivation of the thermodynamic parameters governing interfacial properties in terms of one- and two-solute partition functions.
Specifically, Eqs.~\eqref{def-varphioo} and \eqref{def-veoo} determine the bulk energetic parameter $\ve_\oop$.
Equation~\eqref{def-ko} determines the parameter, $\ko$, that quantifies the intrinsic preference of $o$-solutes for the interface.
Equation~\eqref{def-hoop} determines the curvature, $h_\oop$, of the surface tension.
Finally, given microscopic definitions of $\ve_\oop$, $\ko$, and $h_\oop$, Eqs.~\eqref{eq-ThermoGa} and \eqref{eq-btdtga-xo} give microscopic expressions for the surface excess, $\Gao$, and surface tension, $\ga$, as a function of the composition of the coexisting bulk phase.
Note that while Eq.~\eqref{eq-Gao-xo} gives $\Gao$ as a Taylor series expansion to second order in solute concentration, the rational fraction expression in, e.g., Eq.~\eqref{eq-Gao-two} is expressed in terms of the same parameters, but remains valid over a considerably larger composition regime. 
Consequently, we will adopt the rational fraction expression for $\Gao$ in the following.

\ifjusttheory
\end{document}
\fi

%\newpage
\section{Results and Discussion}
Section~\ref{Sec-Theory} presented a rather general dilute solution theory (DST) for modeling the interfaces of  multicomponent solutions with a dominant solvent ($w$) and multiple dilute cosolute species. %s ($\{o\}$).
% dilute cosolvents upon the interfaces of complex multicomponent solutions.
In order to assess and investigate this DST, we consider the liquid-vapor interface for binary solutions with a single dilute solute species ($o$).
Moreover, we specialize to the case of simple cubic lattices with nearest neighbor interactions.
We first briefly consider this DST for homogeneous solutions and compare with regular solution theory.
We then focus on the properties of liquid-vapor interfaces.
Finally, we briefly employ DST to interpret experimental measurements for solutes that are commonly employed to modulate protein stability.
Throughout this section, we consider a single fixed temperature, $T$, and employ an ``overbar'' to indicate dimensionless quantities that have been scaled by $\bt = 1/\kT$.

\subsection{Homogeneous fluid}
%\subsection{Binary systems: A special case}
%Having provided a general treatment for solutes at interfaces, we now specialize the discussion to binary systems, the focus of this study. To this end, we restate some of the preceding results for a binary mixture of a single solute in solvent. We also examine the dependence of the coefficients introduced in our theory on the microscopic interactions, $u_{ij}$, in the context of our lattice model.

%\subsubsection{Bulk system}

\begin{figure}[h]
	\ifshowfigures
	\includegraphics[scale=0.8]{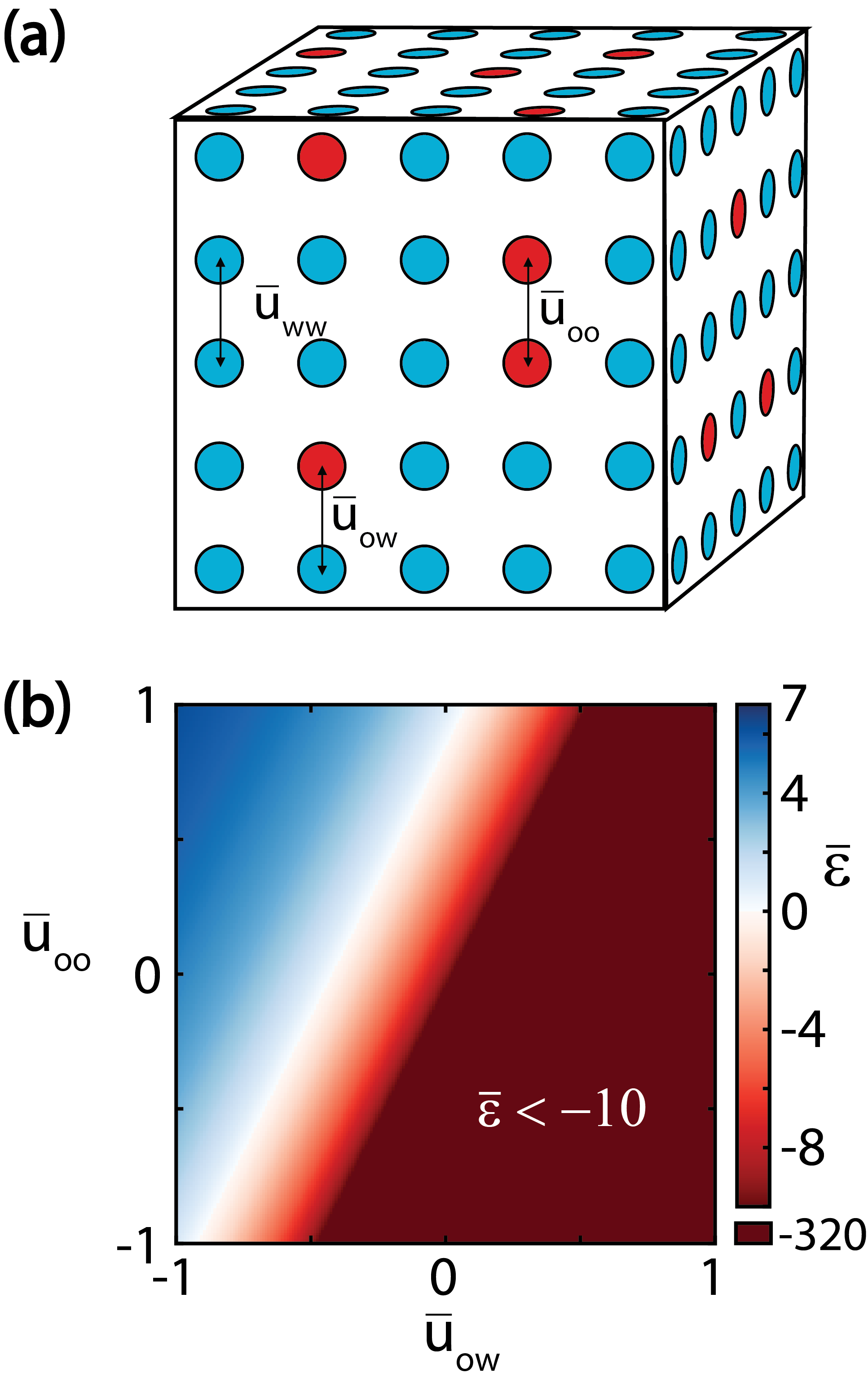}	%Figs/Fig1}
	\fi
	%\centering
	\caption{
	Lattice model for homogeneous binary solution.
	(a) Schematic of cubic lattice model with nearest-neighbor interactions. 
	Blue and red circles represent solvent ($w$) and solute ($o$) molecules, respectively.
	The indicated interaction parameters $\buww$, $\buow$, and $\buoo$ describe the energy of $w$-$w$, $o$-$w$, and $o$-$o$ contacts, respectively. 
	While the reported numerical simulations employ $L$ = 10, the schematic only indicates 5 layers.
	(b) Intensity plot of the thermodynamic interaction parameter, $\ove$, as a function of the microscopic contact energies $\buoo$ and $\buow$.
%Characterizing the bulk coefficient for the binary system. 
%	(a) A representation of the 3-dimensional lattice model considered. Blue circles represent the solvent while the red circles represent the solute. Interactions between species are represented as $u_{ij}$. 
%	(b) The coefficient $\ve_{11}$ depends on the interplay of $u_{ww}, u_{ws}$, and $u_{ss}$. The color bar represents the magnitude of $\ve_{11}$ with a negative value implying a propensity for the solutes to aggregate while a positive value characterizes systems that prefer to stay well solvated. 
}
	\label{fig:heatmaps_bulk}
\end{figure}

We model a homogeneous fluid of $\nw$ solvent molecules and $\no$ solute molecules with a simple three-dimensional cubic lattice of $M=L^3$ sites with $L = 10$, coordination number $z_b = 6$, and periodic boundary conditions in all three directions.
We model interactions with the simple nearest-neighbor contact potential $U(\bR) = \sum_{\llangle i,j \rrangle} u_{t_i t_j}$, where $t_i$ indicates the type of molecule at lattice site $i$ in configuration $\bR$.
We determine the energy scale by fixing the solvent-solvent interaction $\buww = \bt \uww = - 1$.
We consider various solutes by varying the solute-solvent ($\buow$) and solute-solute ($\buoo$) interaction parameters.  
Figure~\ref{fig:heatmaps_bulk}a schematically illustrates the lattice model for the homogeneous binary solution.

Subsection~\ref{SubSec-StatMechHomogeneousSystem} derives the chemical potentials of the homogeneous fluid as a function of solute mole fraction, $x = \no/M$, in the lowest order approximation that treats solute-solute interactions.
The water chemical potential is 
\begin{eqnarray}
\label{eq-muw}
\bmuw 
	& = & 
		\bmuwo - x - \half \ove x^2 + \mO(x^3),	
\end{eqnarray}
where $\bmuwo = \half \zb \buww$ is the chemical potential of pure solvent and $\kT x$ corresponds to the van't Hoff contribution to the osmotic pressure.
The solute chemical potential is 
\begin{eqnarray}
\label{eq-muo}
\bmuo 
	& = &
	 	\bmuost + \ln x + \ove x + \mO(x^2)	,
\end{eqnarray}
where $\bmuost = \zb (\buow - \buww)$  is the energetic cost of introducing a single solute at infinite dilution
%$\bmuost = \zb \dt \bub$ is the energetic cost of introducing a single solute at infinite dilution in terms of $\dt\bub = \buow - \buww$ 
and $\ln x$ corresponds to the ideal mixing entropy for dilute solutes.

The thermodynamic consequences of solute-solute interactions are described by the energetic parameter, $\ove$, which is given by Eq.~\eqref{def-veoo} in terms of the partition functions for a pair of solutes at infinite dilution.
The supplementary material (SM) derives $\ove$ for the cubic lattice with nearest neighbor interactions:
 \begin{equation}
\label{eq:binary_bulk_epsilon}
%\beta \ve =  1 - z_b(e^{\chi } -1 )
%\beta \ve =  1 - z_b[\exp(2\chi/z_b) -1 ]
\ove = 1 - z_b \left[ e^{-\Dtbuoo} - 1 \right]	,
\end{equation}
where $\Dtbuoo = \buoo + \buww - 2 \buow$ is the energy of solute dimerization.
Note that our convention defines attractive interactions by negative energetic parameters, $\bu < 0$. 
Consequently, solute-solute interactions are thermodynamically favorable when $\Dtbuoo, \ove < 0$. 
The energy of solute dimerization becomes increasingly attractive ($\dd\Dtbuoo < 0$) either for increasing solute-solute attraction ($\dd \buoo < 0$) at fixed solute-solvent interaction ($\dd\buow = 0$) or for increasing solute-solvent repulsion ($\dd \buow > 0$) at fixed solute-solute interaction ($\dd\buoo = 0$).

Figure~\ref{fig:heatmaps_bulk}b presents an intensity plot of $\ove$ as a function of the solute-solute ($\buoo$) and solute-solvent ($\buow$) contact energies in the microscopic model. 
According to Eq.~\eqref{eq:binary_bulk_epsilon}, 
%the thermodynamic interaction parameter, $\ove$, 
$\ove$
depends exponentially upon (the negative of) the solute dimerization energy, $\Dtbuoo$.
In the limit of strongly repelling solutes ($\Dtbuoo \to +\infty$), $\ove$ saturates at $\ove \to 1 + \zb = 7$, which corresponds to the solute excluding solvent from its neighboring sites.
Conversely, $\ove$ decreases without bound as $\ove \to - z_b e^{-\Dtbuoo}$ in the limit of strongly attracting solutes ($\Dtbuoo \to -\infty$).
In the limit of weakly interacting solutes ($\Dtbuoo \approx 0$), $\ove \to 1 - 2 \chi$, where $\chi = -\half \zb \Dtbuoo$ is the conventional energetic interaction parameter associated with random mixing and corresponds to the Flory parameter in the Flory-Huggins theory of polymer solutions.\cite{rubinstein2003}
Note that by convention $\chi > 0$ for attractive solutes, which is opposite to the convention we have adopted for $\bu$ and $\ove$.
In the limit of ideal solutions for which solute and solvent molecules interact equivalently ($\Dtbuoo = 0$), $\ove = 1$  due to the excluded volume of molecules on the lattice.

%\twocolumn
\begin{figure}[h]
	\ifshowfigures
	\includegraphics{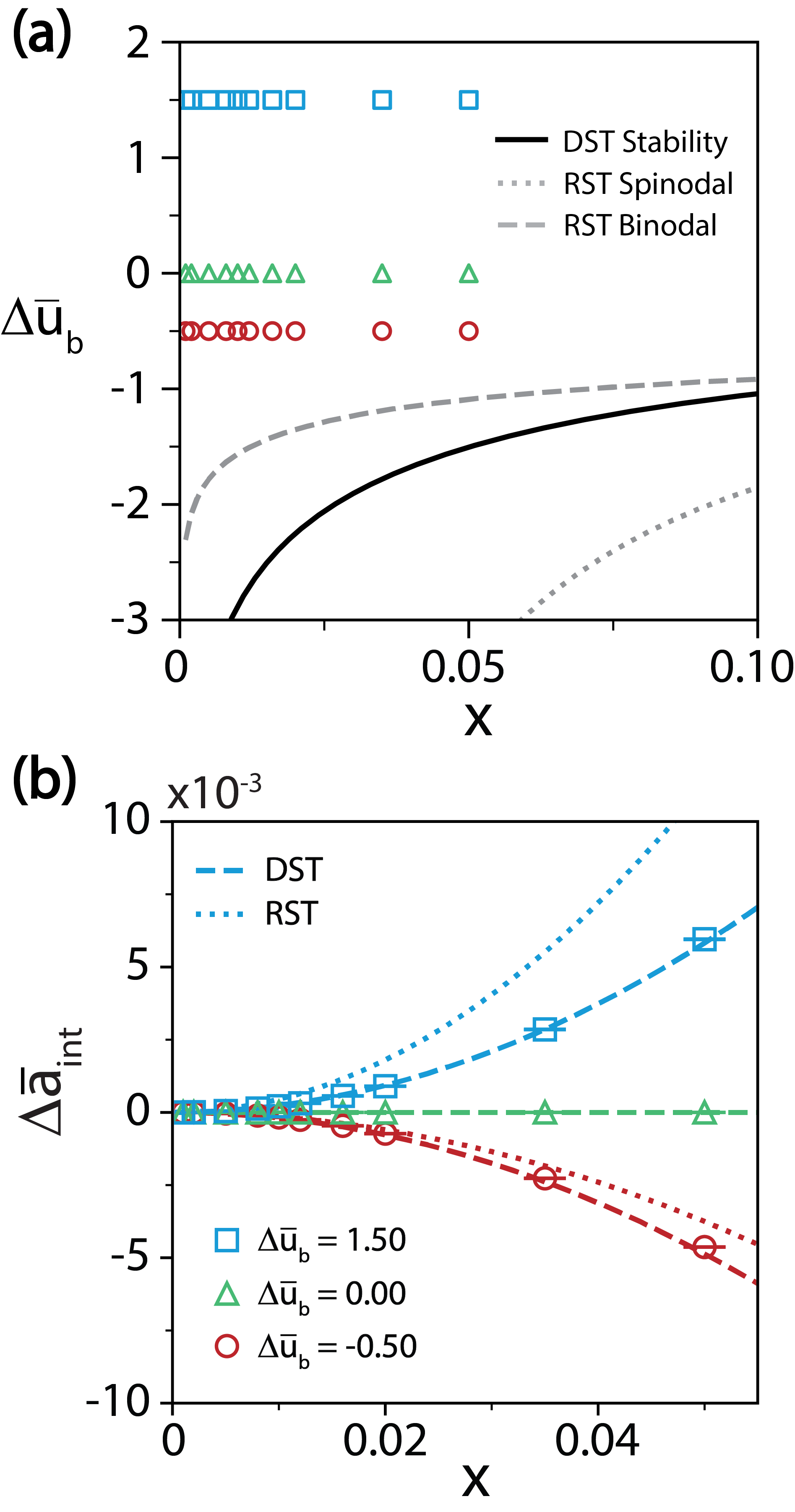}	%Figs/Fig2} 
	\fi
	\centering
	\caption{
	Comparison of dilute solution theory (DST) and regular solution theory (RST) %for the thermodynamic properties of 
	for homogeneous dilute solutions.
	(a) Comparison of predicted phase diagrams in terms of the microscopic dimerization energy, $\Dtbub$, and solute mole fraction, $x$.
 	DST predicts that regions below the solid curve are thermodynamically unstable.
	The dashed and dotted gray curves indicate the binodal and spinodal predicted by RST.
%	The symbols indicate simulated solutions.
	(b) Comparison of theoretical predictions and numerical calculations for the scaled interaction free energy, $\Dt\baint$.
	The dashed and dotted lines present the predictions of DST and RST, respectively. 
	The blue, green, and red curves correspond to solutions with $\Dtbub = $ 1.50, 0.00, and -0.50, respectively. 	
	The symbols present numerically exact calculations of $\Dt\baint$ that are based upon MC simulations of the solutions indicated in panel (a).
	The error bars indicate the simulated uncertainty, which is estimated as the standard deviation across three independent simulations.
%	that are based upon MC simulations and MBAR calculations.	
%	Examining properties of bulk systems 
%	(a) Phase diagram presenting the binodal and spinodal as predicted by our theory (black). Grey lines represent phase boundary predictions according to Flory Huggins theory. Points represent the systems and concentrations simulated via MC, presented in the next panel.
%	(b) Variation of the bulk free energy difference, $\Delta a_{int}$ as a function of solute \conc{} for three different systems with different values of the mixing parameter $\chi$. The dotted lines are predictions from Regular Solution (RS) theory while the dashed lines are predictions from the Semi-Grand Canonical Ensemble (SGCE) theory.
%	Error bars represent the standard deviation across three independent simulations.
}
	\label{fig:bulk_results}
\end{figure}

We define the (scaled) free energy per lattice site, 
$
\ba \equiv M\inv \bt A = \bmuw + x \bmuo
.
$
The present dilute solution theory (DST) gives
%We expand $\ba$ as a :  
\begin{equation}
\label{eq-A}
\ba	
%	 =  
%		\bmuw + x \bmuo 
	= 
		\bmuwo + \left( \bmuost - 1 \right) x + x\ln x + \half \ove x^2 + \mO(x^3)	.
\end{equation}
Because DST treats solutes as a perturbation to pure solvent, it cannot treat concentrated solutions.
Similarly, DST cannot describe dilute solutions under conditions that the solutes begin to aggregate. 
Moreover, the free energy of Eq.~\eqref{eq-A} cannot support two-phase coexistence and becomes unstable when $1 + \ove x < 0$, which is indicated by the solid curve in Fig.~\ref{fig:bulk_results}a.

In order to investigate the influence of solute-solute interactions, we simulate three different classes of solutes under the conditions indicated in Fig.~\ref{fig:bulk_results}a.
The blue symbols correspond to repulsive solutes with $\Dtbuoo = 1.5$; the green symbols correspond to ideal solutes  with $\Dtbuoo = 0$; and the red symbols correspond to attractive solutes with $\Dtbuoo = -0.5$.
If we consider that pure water is approximately 55~M, then $x = 0.02$ corresponds to a solute concentration of approximately 1~M.
In the following, we consider solute concentrations of $x \le 0.05$, which corresponds to $\no \le 50$ solutes on the $M = 10^3$ lattice sites.
Figure~\ref{fig:bulk_results}a indicates that each of these solutions should remain stable under these conditions.

The SM demonstrates that the present DST very accurately describes $\ba$ for these solutions.
However, under these dilute conditions $\ba$ is largely determined by the first three terms in Eq.~\eqref{eq-A}, which describe bulk solvent, isolated solutes, and ideal mixing, respectively. 
%infinitely dilute solutions and do not reflect solute-solute interactions.
In order to eliminate these trivial contributions from infinitely dilute solutes, we define a reference system in which the solute and solvent molecules interact equivalently (i.e., $\Dt\bu_{b;\rm Ref} = 0$). 
We then define the interaction free energy as, $\Dt\baint = \ba - \ba_{\rm Ref} - x \muost$, which eliminates the infinitely dilute contributions and focuses on solute-solute interactions. 
%, between the free energy of the system and the free energy of the reference system.

The dashed curves in Fig.~\ref{fig:bulk_results}b present the predictions of DST for the interaction free energy, $\Dt\baint,$ as a function of solute mole fraction, $x$, for the repulsive, ideal, and attractive solutes of Fig.~\ref{fig:bulk_results}a.
As expected, the interaction free energy vanishes for ideal solutes, but quadratically increases and decreases for repulsive and attractive solutes, respectively.
The symbols in Fig.~\ref{fig:bulk_results}b present numerically exact calculations of $\Dt\baint$ based upon analyzing Monte Carlo (MC) simulations via the multistate Bennett Acceptance Ratio (MBAR) method.
Although DST is completely determined by 1- and 2-solute partition functions and only treats interactions to lowest order, it accurately predicts the numerically exact calculations for solute concentrations up to $x = 0.05$.

Finally, before considering interfacial phenomena, it is worth briefly comparing DST with regular solution theory (RST).\cite{hill1986,rubinstein2003}
The scaled free energy for RST is 
\begin{equation}
\ba_{\rm RST} = \bmuwo + \bmu_o^\phi + x \ln x + \xw \ln \xw + \chi x \xw	,
\end{equation}
where $\bmu_o^\phi = \half\zb \buoo$ is the chemical potential of bulk solute, $\xw = 1 - x$ is the mole fraction of solvent, and $\chi = -\half \zb \Dtbuoo$ is again the Flory interaction parameter.
The RST free energy is usually derived by exactly treating the combinatoric entropy of mixing, while adopting the Bragg-Williams (i.e., random mixing) approximation for the interaction energy. 
RST exactly describes ideal solutions and correctly reproduces the infinite dilution contributions to the free energy.
Because it exactly treats the mixing entropy, RST also provides a qualitatively reasonable description of concentrated solutions and is particularly useful for modeling phase coexistence.

The dotted and dashed curves in Fig.~\ref{fig:bulk_results}a present the binodal and spinodal curves predicted by RST. 
The dotted curves in Fig.~\ref{fig:bulk_results}b present the predictions of RST for the interaction free energy, $\Dt\baint$. 
%While RST exactly describes ideal solutions and correctly reproduces the infinite dilution contributions to the free energy, 
%the dotted curves in Fig.~\ref{fig:bulk_results}b demonstrate that
As expected for a mean field theory, RST systematically overestimates the interaction free energy.
In particular, RST significantly over-estimates the likelihood of repulsive solutes contacting.
Conversely, RST underestimates the likelihood of attractive solutes contacting.
In comparison, DST provides a significantly more accurate treatment of solute-solute interactions for these dilute solutions. %$x \le 0.05$.

\subsection{Inhomogeneous system}
\label{SubSec-ResultsInhomogIntro}
\begin{figure}[h]
	\ifshowfigures
	\includegraphics[scale=0.8]{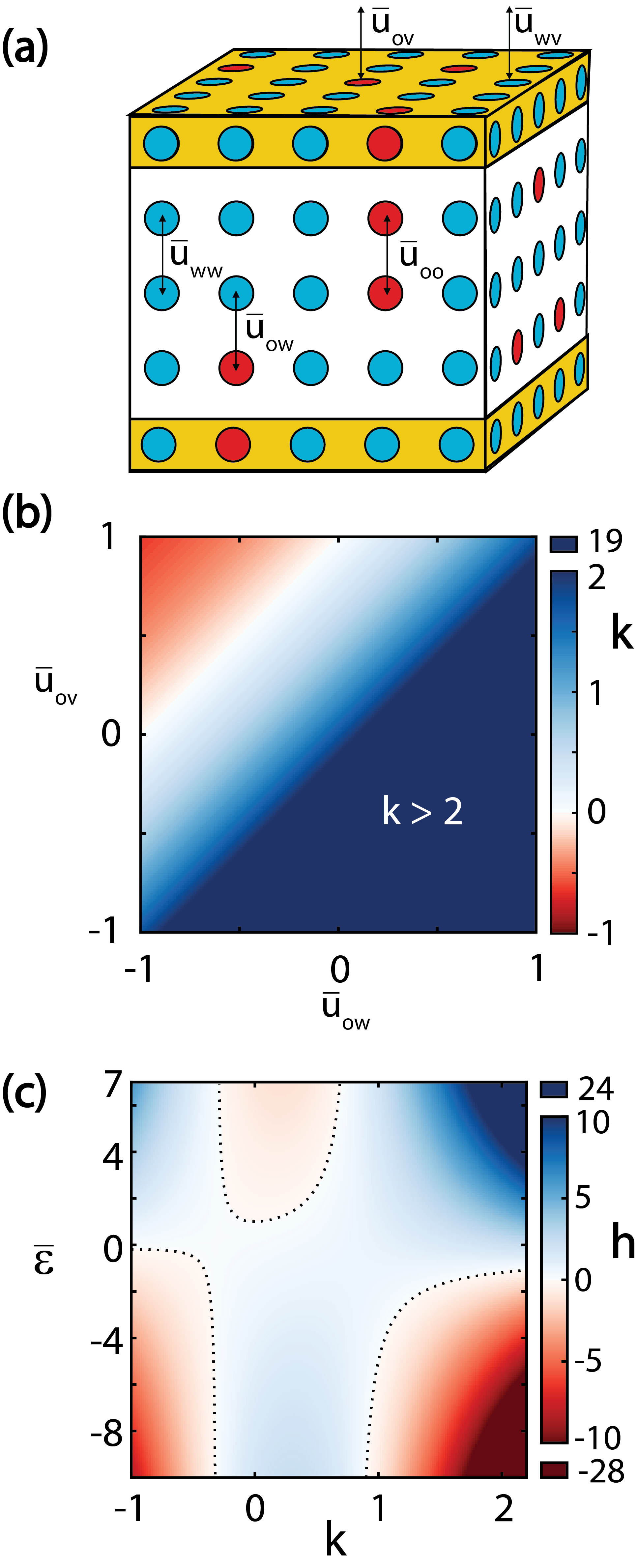}	%Figs/Fig3}
	\fi
	\centering
	\caption{
	Lattice model of inhomogeneous binary system.
	(a) Schematic of cubic lattice model with nearest-neighbor interactions. 
	Blue and red circles represent solvent ($w$) and solute ($o$) molecules, respectively, while the yellow region indicates the interfacial sites, $\cS$.
	The parameters $\buwv \equiv 0$ and $\buov$ describe the microscopic energetics of solvent and solute molecules, respectively, in the interfacial region. 
%	The interaction parameters $\buww$, $\buow$, and $\buoo$ describe the energy of $w$-$w$, $o$-$w$, and $o$-$o$ contacts, respectively. 
	While the reported numerical simulations employ $L$ = 10, the schematic only indicates 5 layers.
	(b) Intensity plot of the parameter, $k$, that describes the intrinsic interfacial preference of solutes as a function of the microscopic energies $\buow$ and $\buov$.	
	(c) Intensity plot of the parameter, $h = \ptl^2 \bga/\ptl x^2$, that describes the curvature of the surface tension as a function of the thermodynamic parameters $k$ and $\ove$.
	Dotted lines indicate where $h = 0$.
%	Characterizing the surface tension coefficients for a binary system. 
%	(a) A representation of the interfacial lattice. The unshaded region represents the bulk phase, analogous to Fig. \ref{fig:heatmaps_bulk}. The yellow regions represent the interface and species in this region interact with a hypothetical vapor phase with interactions represented as $u_{iv}$. 
%	(b) The heatmap represents the dependence of $k$, the intrinsic preference of the solute for the interface, as a function of the water-solute and the solute-vapor interactions. The colors show the predicted values of the coefficient.
%	(c) The heatmap represents the dependence of $h$, the curvature of the surface tension, which is completely defined by the interfacial preference ($k$) and the bulk interaction parameter ($\chi$). The colors show the predicted values of the coefficient. 
}
	\label{fig:heatmaps_surften}
\end{figure}

We model an inhomogeneous system of $\Nw$ solvent molecules and $\No$ solute molecules on a corresponding three-dimensional cubic lattice of $M = L^3$ sites with $L = 10$. 
The $M_i = 2 L^2$ lattice sites in the top ($z=1$) and bottom ($z=L$) layers define an interfacial region, $\cS$, in which molecules experience an external field.
The remaining $M - M_i$ sites in the intervening $L-2$ layers then describe a bulk liquid region between the two interfaces.
The potential is $U(\bR) = \sum_{\llangle i,j \rrangle} u_{t_i t_j} + \sum_{i\in\cS} u_{t_i v}$, where the first sum describes nearest neighbor interactions with periodic boundary conditions in the x- and y- directions, while the second sum describes the interaction of interfacial molecules with the external field.  
We mimic the hydrophobic air-water interface by defining $\buwv = 0$.
We vary the parameter $\buov$ to modulate the preference of the solute for the interface. 
With the exception of Subsection~\ref{SubSec-ResultsPolarizn}, we neglect polarization effects and assume that intermolecular interactions are the same throughout the inhomogeneous system. 
%However, we briefly consider polarization effects in Subsection~\ref{SubSec-ResultsPolarizn}.
Figure~\ref{fig:heatmaps_surften}a schematically illustrates the model.

Subsection~\ref{SubSec-StatMech} derived analytic expressions for the surface tension and surface excess in the lowest order approximation that treats solute-solute interactions.
The (scaled) surface tension, $\bga = \bt\ga$, may be expressed 
\begin{equation} 
\label{eq:binary_lattice_surften_fit}
\bga = 
	\bga^\phi - k x + \half h x^2 + \mathcal{O}(x^3)		,
\end{equation}
where $x$ is the mole fraction of solute in the coexisting bulk region, while $\bga^\phi = \bt\ga^\phi$ is the (scaled) surface tension of pure solvent.
For the simple lattice model $\bga^\phi = \half \left(\zi - \zb\right)\buww + \buwv = \half$, where $\zi = 5$ is the number of intermolecular interactions experienced by interfacial molecules.
According to Eq.~\eqref{eq-Gao-one}, the surface excess may be approximated 
\begin{equation}
\label{eq:binary_theory_pref1} 
\Ga =
	x \rdbracket{\frac{k - h x}{1 + \ove x} } 	,
\end{equation}
where $\ove$ is the bulk interaction parameter given by Eq.~\eqref{eq:binary_bulk_epsilon}.

The parameter $k = \left. \ptl \Ga / \ptl x \right|_0 = - \left. \ptl \bga / \ptl x \right|_0$ describes the ``intrinsic preference'' of solutes for the interface.
(Here and elsewhere $\left. \right|_0$ indicates that the partial derivatives are evaluated for $x = 0$.)
While Eq.~\eqref{def-ko} gives a general expression for $k$ in terms of effective one-solute partition functions, the SM explicitly derives $k$ for the simple nearest-neighbor lattice model:
\begin{equation}
\label{eq-k}
k =  \exp\left[- \Dtbui \right] - 1	,
%. \beta( \delta u_b - \delta u_i)] -1	, 
%k =  \exp [\beta( \delta u_b - \delta u_i)] -1
\end{equation}
where 
%$\Dtbui= (\buov - \buwv) -  (\buow - \buww)$ 
$\Dtbui= \buov + \buww  - ( \buwv + \buow ) $ 
is the microscopic interfacial energy of the solute, i.e., the energetic cost of moving an infinitely dilute solute from the bulk to the interface.
This interfacial energy, $\Dtbui$, becomes more favorable ($\dd\Dtbui < 0$) when either the solute-interface energy becomes more favorable ($\dd\buov < 0$) or the solute-solvent interaction becomes more repulsive ($\dd\buow > 0$).
Figure~\ref{fig:heatmaps_surften}b presents an intensity plot of the intrinsic interfacial preference, $k$, as a function of the microscopic solute-interface ($\buov$) and solute-solvent ($\buow$) energies.
For molecules with favorable interfacial energy ($\Dtbui < 0$), $k$ exponentially increases with $|\Dtbui|$ and, in particular, increases without bound as $\Dtbui \to -\infty$.
Conversely, for molecules with very unfavorable interfacial energy ($\Dtbui \to +\infty$), $k$ saturates at -1.
%Conversely, for strong depletants ($\Dtbui \to +\infty$), $k$ saturates at -1.
%For strong depletants ($\Dtbui \to +\infty$), $k$ saturates at -1.
%Conversely, for strong surfactants ($\Dtbui \to - \infty$), $k$ increases without bound.
For molecules with minimal interfacial energy ($\Dtbui \approx 0$), $k \to - \Dtbui$.

\newcommand{\tk}{\tilde{k}}

Equation~\eqref{eq:binary_theory_pref1} clearly demonstrates that interactions with other solutes can impact the interfacial preference.
Accordingly, we define the composition-dependent ``effective preference'' of solutes for the interface by 
\begin{equation}
\label{def-tk}
\tk(x) \equiv \dd \Ga(x) / \dd x = k - 2 (h + k\ove) x + \mO(x^2)	.
\end{equation}
The effective preference, $\tk(x)$, quantifies the tendency of additional solutes to partition to the interface when the mole fraction of the bulk region is $x$. 
In the limit $x\to 0$, $\tk(x)$ reduces to the intrinsic interfacial preference, $k$.
The effective preference varies in a manner that reflects not only $h$, but also $k$ and the bulk solute-solute interaction energy, $\ove$.  
Equation~\eqref{def-tk} suggests that solute-solute interactions can convert surfactants ($k > 0$) to effective depletants if $h + k\ove > 0$. 
Conversely, these interactions can convert depletants ($k < 0$) to effective surfactants if $ h + k\ove < 0$.

The parameter, $h$ is of particular interest for describing the influence of solute-solute interactions upon interfaces.
%According to Eq.~\eqref{eq:binary_lattice_surften_fit}, $h$ determines the curvature of $\ga$ as a function of solute concentration.
%Moreover, according to Eq.~\eqref{eq:binary_theory_pref1}, $h$ quantifies the effect by which solute-solute interactions can convert depletants ($\Ga < 0$) to surfactants ($\Ga > 0$) and vice versa.
Equation~\eqref{def-hoop} gives a general expression for $h_\oop = \left. \ptl^2 \bga/\ptl \xo \ptl \xop \right|_0$ in multicomponent solutions in terms of microscopic partition functions for two solutes at infinite dilution.
For binary solutions, this expression reduces to 
\begin{equation}
\label{eq:binary_beta}
h 
	=
		 \eti^{-1}\sqbracket{r^2 \ove_t + \ove( 1 - 2r)  }	.
\end{equation}
In Eq.~\eqref{eq:binary_beta} $\eti \equiv M_i/M$ is the surface/volume ratio of the inhomogeneous system; 
$r = 1 + \eti k$ is defined in Eq.~\eqref{eq:r_k} as a ratio of one-solute partition functions for the inhomogeneous and homogeneous systems; 
%$r$ is defined in Eq.~\eqref{eq:r_k} as a ratio of one-solute partition functions for the homogeneous and inhomogeneous systems that may be expressed $r = 1 + \eti k$; 
and $\ove_t$ is an energetic parameter analogous to $\ove$ that describes the free energy of the inhomogeneous system as a function of $x_t \equiv \No / M$.
While Eq.~\eqref{eq:binary_beta} clearly reflects finite size effects, one expects that in the thermodynamic limit $\eti \to 0$ and $\ove_t \to \ove + \eti \dt\ove_i + \mO(\eti^2)$, where $\dt\ove_i$ is independent of system size and describes the influence of interactions at the interface.
In this limit, $h \to \dt\ove_i $.
The SM analyzes these finite size effects in greater detail and, in particular, demonstrates that $h$ becomes independent of $L$ for this nearest-neighbor lattice model when $L \ge 3$.

In the absence of polarization effects, the lattice model for the inhomogeneous system depends upon three energetic parameters that describe the interactions of solute molecules with the solvent ($\buow$), with other solutes ($\buoo$), and with the external field describing the liquid-vapor interface ($\buov$).
Since we have treated $\buov$ as independent variable, we expect that $\bga$ should be independent of $\bmuost = \zb\left( \buow - \buww \right)$. 
Consequently, we anticipate that $h$ should depend upon only two energetic parameters:
(1) the dimerization energy, $\Dtbub = \buoo + \buww - 2 \buow$, which determines $\ove$; 
and 
(2) the interfacial energy,
%$\Dtbui = (\uov - \uow) - (\uwv - \uww)$, 
$\Dtbui = \buov + \buww  - (\buwv + \buow)$, 
which determines $k$.
While the analytic expression for $h$ is somewhat cumbersome, the SM indicates that, for this simple lattice model, $h$ is indeed completely determined by $\Dtbub$ and $\Dtbui$.
% these two parameters.
%Thus, the curvature of the surface tension is completely determined by the bulk solute-solute interaction and the intrinsic preference of the solute for the interface.

Figure~\ref{fig:heatmaps_surften}c presents an intensity plot of $h$ as a function of $k$ and $\ove$.
The dotted curves indicate solute parameters for which $h = 0$.
For these solutes the surface tension will vary linearly with solute concentration up to relatively high concentrations. 
One might expect that $h = 0$ only when solutes interact weakly.
However, Fig.~\ref{fig:heatmaps_surften}c suggests that linear surface tension variations will be observed for a wide range of solutes, including strongly interacting solutes.
Moreover, this intensity plot demonstrates several interesting regimes.

In the case of depletants ($k \to -1$), $h > 0$ for repelling solutes ($\ove > 0$) and $h < 0$ for attracting solutes ($\ove < 0$).
For $k \approx -1$, the effective interfacial preference, $\tk(x) \approx k - 2 (h - \ove)x \approx k$ for all $x$ because $(h-\ove)x$ tends to be rather small.
Thus, the interfacial preference of strong depletants is only weakly influenced by interactions with other depletants.

In the case of surfactants ($ k \gtrsim 1 $), $h$ and $\ove$ again tend to have the same sign.
However, because $k$ is positive in this case, $\dd\tk/\dd x = - 2(h + k \ove)$ can potentially become significant.
In particular, the presence of attracting solutes ($h, \ove < 0$) can increase the effective interfacial preference ($\dd\tk/\dd x > 0$) of surfactants. 
%, i.e., attractive interactions ($\ove < 0$) can enhance the effective interfacial preference of surfactants for the interface. 
Thus, attractive interactions can potentially enhance the surface activity of surfactants.
Conversely, the presence of repulsive solutes ($h, \ove > 0$) can reduce the effective interfacial preference ($\dd\tk/\dd x < 0$) of surfactants. 
In principle, repulsive interactions can potentially convert surfactants into effective depletants.

Figure~\ref{fig:heatmaps_surften}c also demonstrates an intermediate regime of ``surface-neutral'' solutes with $-0.4 \lesssim k \lesssim 1$.
In contrast to depletants and surfactants, $h$ and $\ove$ tend to be anti-correlated for surface-neutral molecules. 
In this regime, $h$ is relatively small for all the solutes that we consider. 
%Consequently, the surface tension should vary approximately linearly over a relatively wide solute concentration range.
Nevertheless, because $k \approx 0$, $\dd\tk/\dd x \approx - 2h$.
This suggests that repulsive solute-solute interactions ($\ove > 0, h< 0$) can convert weak depletants ($k = -\dt$) into effective weak surfactants ($\tk > 0$). %= +\tilde\dt$).
Conversely, attractive solute-solute interactions ($\ove < 0, h> 0$) can convert weak surfactants ($k = +\dt$) into effective weak depletants ($\tk < 0$).
%(Here $\dt$ and $\tilde\dt$ indicate small positive numbers.)
(Here $\dt$ indicates a small positive number.)

In the following, we employ canonical MC simulations to investigate these effects. 
%We define $x_t = \No / M$ as the fixed mole fraction of solutes throughout the entire system.
While the total number of solutes, $\No$, is fixed in these simulations, the number of solutes in the bulk, 
%$\widetilde{N}_{o|b}(\bR)$, 
$N_{o|b}(\bR)$, 
and interfacial regions, 
%$\widetilde{N}_{o|i}(\bR)$, 
$N_{o|i}(\bR)$, 
fluctuate as a function of the system configuration, $\bR$.
We determine the composition of the solution by the average number of solutes in the bulk region, 
%$x = M_b\inv \llangle \widetilde{N}_{o|b}(\bR) \rrangle$ 
$x = M_b\inv \llangle N_{o|b}(\bR) \rrangle$. 
%as the average mole fraction of solutes in the bulk region. 
In general, the composition of the bulk and interfacial regions will differ due to the presence of the interface, i.e., $x \neq \No/M$.
We determine the solute surface excess by $\Ga = M_i\inv \left( \No - x M \right) $.
We determine the surface tension, $\ga$, by employing MBAR to determine the total free energy, $A_t$, of the inhomogeneous system and then subtracting off the bulk contributions.
Section~\ref{Sec-Methods} and the SM describe our computational methods in detail.

%% THIS IS THE END OF VARUN'S ORIGINAL

\subsection{Intrinsic surface preferences} 
\begin{figure}[h]
	\ifshowfigures
	\includegraphics[scale=0.7]{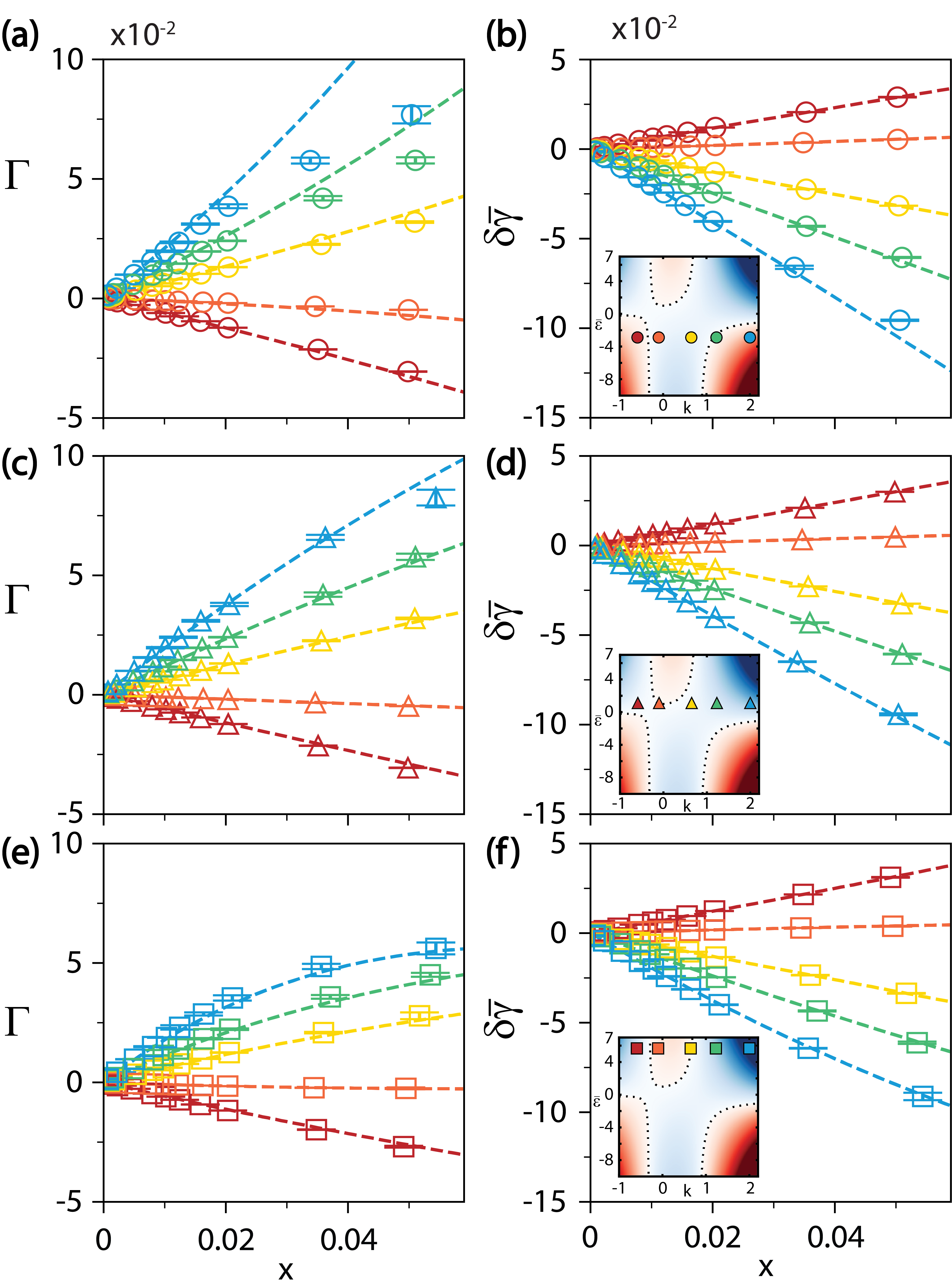}	%Figs/Fig4}
	\fi
	\centering
	\caption{
	Interfacial properties of lattice models.
	The left panels (a,c,e) present the surface excess, $\Ga$, while the right panels (b,d,f) present the variation in the surface tension, $\dt\bga$, as a function of the composition, $x$, of the coexisting bulk phase.
	The top row (panels a and b) presents results for attractive solutes with $\Dtbub = -0.5$; 
	the middle row (panels c and d) presents results for ideal solutes with $\Dtbub = 0$;
	and 
	the bottom row (panels e and f) presents results for repulsive solutes with $\Dtbub = 1.5$. 
	In each case, the various colors indicate various intrinsic preferences for the interface, $k$.
	Red, orange, yellow, green, and blue curves correspond to $k = $ -0.60, -0.10, 0.65, 1.23, and 2.00, respectively. 
	The insets of the right panels indicate the simulated models on the intensity plot for the $h$ parameter.
	In each panel, the dashed curves indicate the analytic predictions of DST, while the symbols present numerical estimates based upon MC simulations and MBAR calculations.
The error bars indicate the estimated uncertainty in these numerical calculations based upon the standard deviation in the results of three independent MC simulations.
%
%	(a) Preferential interaction as a function of \osm{} \conc{} for a set of self-interacting solutes with fixed $\chi = 1.5$ and variable interfacial preference.
%	(b) Surface tension for the same systems as in (a). The inset represents the space occupied by these systems on the $h$ heat map.
%	(c) Preferential interaction as a function of \osm{} \conc{} for a set of perfect systems with fixed $\chi = 0.0$ and variable interfacial preference.
%	(d) Surface tension for the same systems as in (c). The inset represents the space occupied by these systems on the $h$ heat map.
%	(e) Preferential interaction as a function of \osm{} \conc{} for a set of repulsive solutes with fixed $\chi = -4.5$ and variable interfacial preference.
%	(f) Surface tension for the same systems as in (e). The inset represents the space occupied by these systems on the $h$ heat map.
%	Error bars represent the standard deviation across three independent simulations.  
}
	\label{fig:interfacial}
\end{figure}

We first consider the influence of the intrinsic interfacial preference, $k$, upon the surface excess, $\Ga$, and (scaled) surface tension, $\bdtga = \bt( \ga - \ga^\phi)$.
The top, middle, and bottom rows of Fig.~\ref{fig:interfacial} correspond to the three representative classes of solutes from Fig.~\ref{fig:bulk_results}a with net attractive, vanishing, and repulsive solute-solute interactions, respectively. 
For each class of solutes with fixed solute-solute interaction, $\Dtbub$, we systematically vary the intrinsic interfacial preference of the solute from $k \approx -0.6$ (red) to $k = +2$ (blue).
In each panel, the dashed curves present predictions of the analytic DST, while the symbols present the results of MC simulations.  
%The insets of the right column indicate each solute on the intensity plot for $h$.

One expects that solutes with a strong intrinsic preference for the interface ($k = +2$, blue) will accumulate at the interface such that $\Ga$ will increase and $\bga$ will decrease with increasing solute concentration.
As $k$ decreases at fixed composition, one expects that $\Ga$ will systematically decrease and $\ga$ will increase.
One expects that solutes with a strong intrinsic preference for the bulk ($k = -0.6$, red) will be depleted from the interface such that $\Ga$ will decrease and $\bdtga$ will increase with increasing solute concentration.
A cursory glance at Fig.~\ref{fig:interfacial} demonstrates that the analytic DST and MC simulations are both consistent with these basic considerations.
However, these considerations do not account for the influence of solute-solute interactions.
%, which are quantified by $h$.

\begin{comment}
One expects that solutes with a strong preference for the interface will adsorb at the interface such that $\Ga$ will increase and $\bdtga$ will decrease with increasing solute concentration.
Conversely, one expects that solutes with a strong preference for the bulk will be depleted from the interface such that $\Ga$ will decrease and $\bdtga$ will increase with increasing solute concentration.
\end{comment}

The top row of Fig.~\ref{fig:interfacial} considers attractive solutes with $\Dtbub = -0.5$. 
For each fixed surface preference, $k$, the surface tension varies nearly linearly with solute concentration.
DST predictions and numerical simulations agree almost quantitatively in all cases except for strongly surface active molecules at very high concentration $x \approx 0.05$.
In this case, DST slightly overestimates the negative curvature of the surface tension and, consequently, slightly underestimates $\bga$ at the highest concentration.
The surface excess also tends to vary quite linearly with solute concentration.
%However, 
For the strongest surfactants ($k = $ +1.23, +2, green and blue) DST predicts that the effective interfacial preference, $\tk$, should increase with solute concentration and, therefore, that $\Ga$ should demonstrate positive curvature.
When compared to the MC simulations, DST significantly overestimates $\Ga$ for strong surfactants when $x > 0.03$.
We anticipate that DST performs relatively poorly in this case because the dilute solution expansion begins to fail as many solutes start aggregating at the interface.
%significantly overestimates the positive curvature of  
%However, at relatively high concentrations, $x > 0.03$, solute-solute interactions slightly reduce the surface excess of highly surface active molecules with $k \ge 1$.
%This effect is not captured by the analytic model since the lowest order expansion begins to fail when many solutes start aggregating at the interface.
Aside from this discrepancy for relatively concentrated solutions of strong surfactants, the analytic theory quite accurately predicts $\Ga$ for these attractive solutes, especially for depletants and weak surfactants.

The middle row of Fig.~\ref{fig:interfacial} considers ideal solutes for which $\Dtbub = 0$ (i.e., $\ove = 1$).
The surface tension varies almost linearly with concentration in this case, although for strong surfactants, $\bga$ demonstrates slight positive curvature.
The surface excess, $\Ga$, varies approximately linearly for $k < 1$.
However, $\Ga$ demonstrates noticeable negative curvature for surfactants with $k \ge 1$. 
In the case of ideal solutes, DST predicts both $\Ga$ and $\bga$ with almost quantitative accuracy for all $ x \le 0.05$.

Finally, the bottom row of Fig.~\ref{fig:interfacial} considers repulsive solutes with $\Dtbub = 1.5$.
The surface tension varies quite linearly for $k < 1$, but demonstrates significant positive curvature for surfactants with $k \ge 1$.
Moreover, the surface excess demonstrates pronounced negative curvature for strong surfactants.
%It is interesting that these nonlinearities become more pronounced for solutes with greater surface preference because solute-solute interactions play a greater role as more solutes accumulate at the surface.
%In this case, repulsive solute-solute interactions dramatically suppress the accumulation of surfactants at the interface.
Repulsive solute-solute interactions dramatically suppress the accumulation of surfactants at the interface.
In this case of repulsive solutes, DST predicts both $\Ga$ and $\bga$ for all solutions with nearly quantitative accuracy.

%%% THIS IS THE END OF VARUNS' ORIGINAL

\subsection{Transitions in surface activity} %Interaction-driven surface activity transitions}

\begin{figure}[h]
	\ifshowfigures
	\includegraphics[scale=0.8]{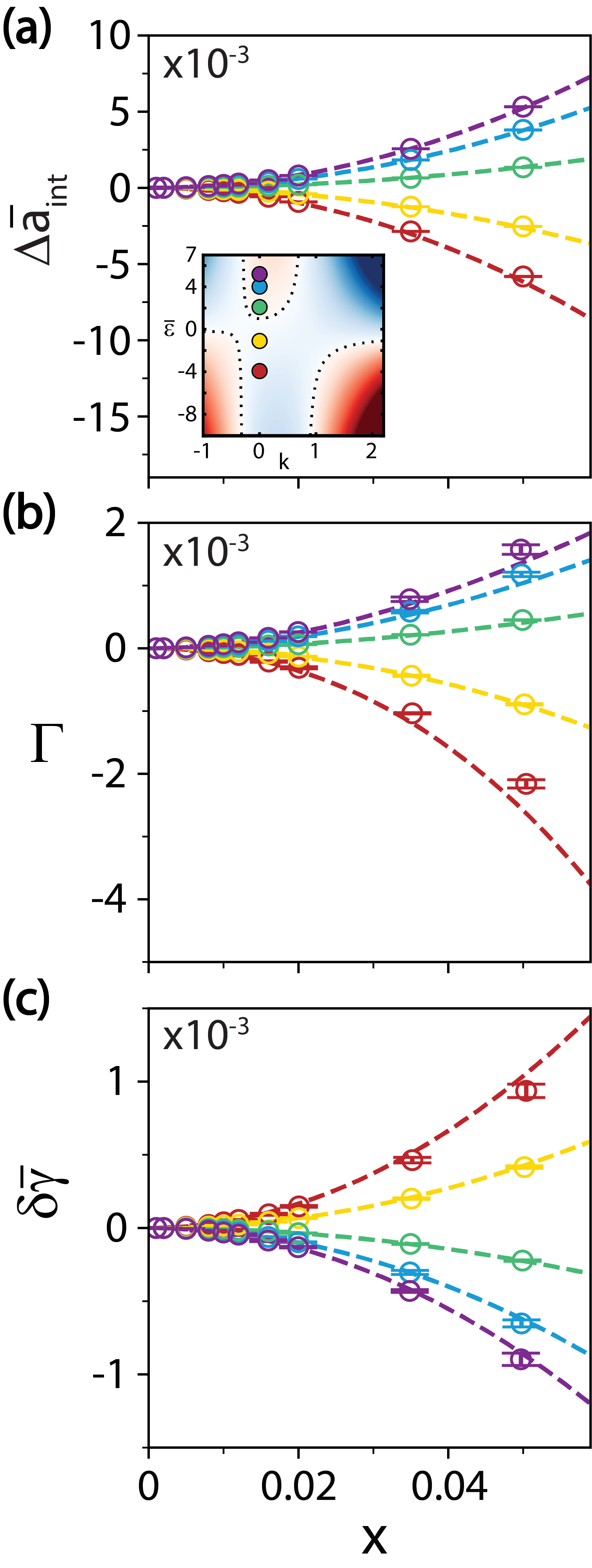}	%Figs/Fig5_inset}
	\fi
	\centering
	\caption{
Bulk and interfacial properties of surface-neutral molecules with $k = 0.$
Panels a, b, and c present the bulk interaction free energy, $\Dt\baint$, the surface excess, $\Ga$, and the surface tension variation, $\dt\bga$, as a function of the solution composition, $x$.
The inset of panel a indicates the simulated systems on the $h$ intensity plot.
Red, yellow, green, blue, and purple present results for solutes with $\Dtbub = $ -0.6, -0.3, 0.2, 0.7, and 1.2, respectively. 
In each panel, the dashed curves indicate the analytic predictions of DST, while the symbols present numerical estimates based upon MC simulations and MBAR calculations.
The error bars indicate the estimated uncertainty in these numerical calculations based upon the standard deviation in the results of three independent MC simulations.
%	Systems with no intrinsic interfacial preference:
%    (a) Bulk free energy differences of the systems compared to a reference system. Each point in the inset represents different systems with fixed interfacial preference ($k = 0$) and varying bulk interactions ($-3.6 < \chi < 2.1$). Colors on the heatmap represent the curvature of the surface tension, $h$.
%    (b) Surface tension variations as a function of concentration for the different systems. 
%    (c) The preferential interaction coefficient plotted as a function of concentration. In panels The dashed lines represent predictions from the theory while the points represent values from simulations (see Methods for details on the calculation). 
% Error bars represent the standard deviation across three independent simulations.
}
	\label{fig:k_1}
\end{figure}

%\twocolumn

In order to more carefully investigate the influence of solute-solute interactions upon interfacial properties, we next consider a class of surface-neutral molecules with no intrinsic preference for the interface, i.e., $k = 0$.
We vary the solute-solute interaction energy from quite attractive $\Dtbub = -0.6$ (red) to rather repulsive $\Dtbub = +1.2$ (purple).
Figure~\ref{fig:k_1}a presents the bulk interaction free energy, $\Dt\baint$, for these solutes. 
%, while the inset indicates these solutes on the intensity map for $h$.
The inset of Fig.~\ref{fig:k_1}a indicates that the curvature, $h$, of the surface tension flips sign as $\ove$ increases.
In particular, $h > 0$ for strongly attracting solutes, while $h < 0$ for strongly repelling solutes.

Figures~\ref{fig:k_1}b and \ref{fig:k_1}c present the surface excess, $\Ga$, and surface tension variation, $\dt\bga$, for these surface-neutral solutes.
Because $k = 0$, $\Ga$ and $\bdtga$ are roughly an order of magnitude smaller for surface-neutral solutes than for the more typical solutes considered in Fig.~\ref{fig:interfacial}.
According to Subsection~\ref{SubSec-ResultsInhomogIntro}, DST predicts that the surface tension should vary as $\bdtga = \half h x^2 + \mO(x^3)$, the surface excess should vary as $\Ga = - h x^2 + \mO(x^3)$, and the effective interfacial preference should vary as $\tk = - 2 h x + \mO(x^2)$.
Figures~\ref{fig:k_1}b and \ref{fig:k_1}c demonstrate that these theoretical predictions match the numerical simulations quite accurately.
In particular, surface-neutral solutes with attractive interactions act as effective depletants reducing $\Ga$ and increasing $\bga$.
Conversely, surface-neutral solutes with repulsive interactions act as effective surfactants increasing $\Ga$ and decreasing $\bga$.

%% THIS WAS MY ORIGINAL

\begin{figure}[h]
	\ifshowfigures
	\includegraphics[scale=0.8]{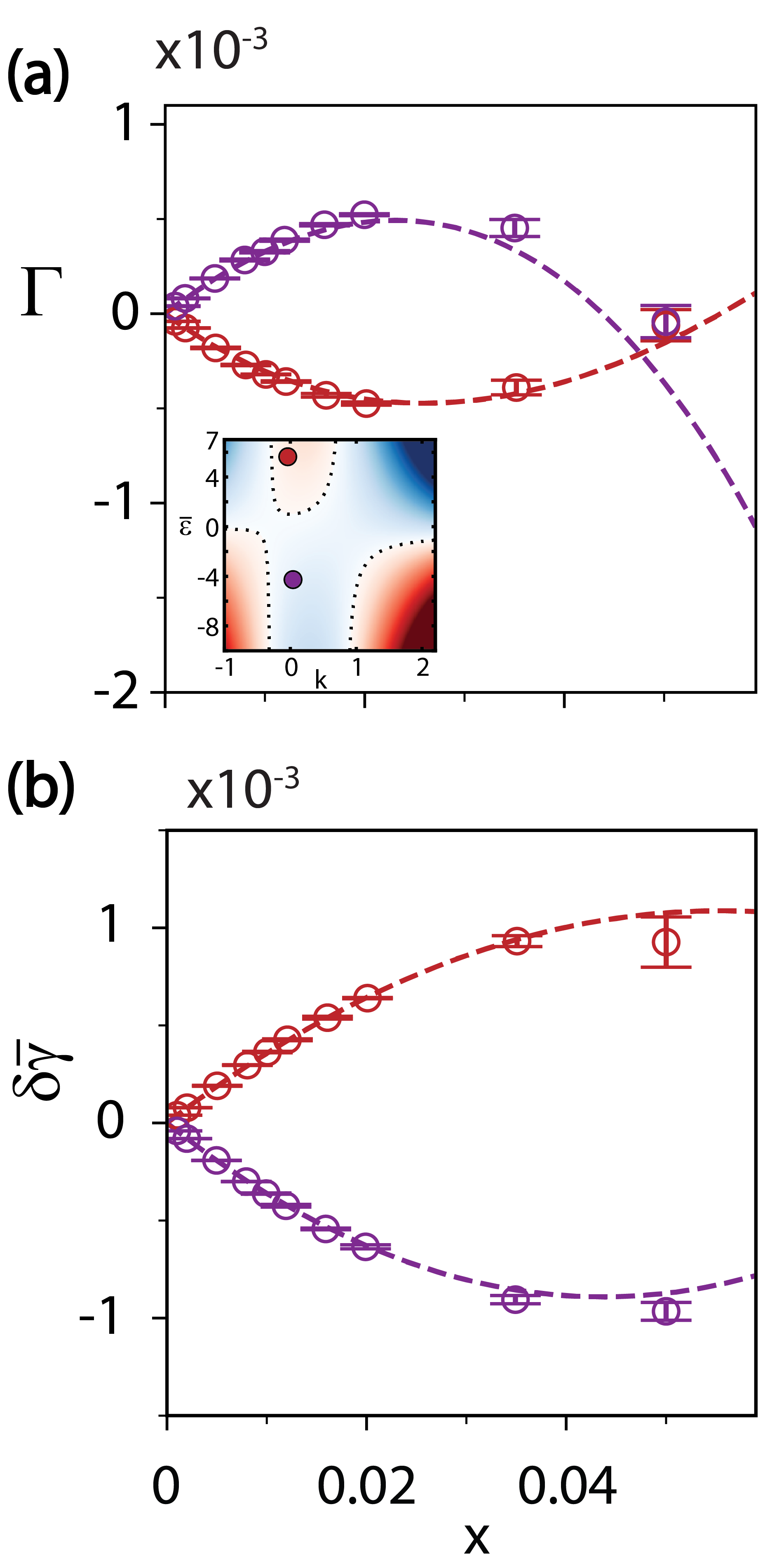}	%Figs/Fig6_inset}
	\fi
	\centering
	\caption{ 
	Surface activity transitions.
	Panels a and b present the surface excess, $\Ga$, and surface tension variation, $\dt\bga$, respectively, as a function of solute concentration, $x$.
	The red curves correspond to $\Dtbub = $ +1.48 and $\Dtbui = $ +0.04, while the purple curves correspond to $\Dtbub = $ -0.63 and $\Dtbui = $ -0.04. 
	The inset of panel a indicates the two simulated systems on the $h$ intensity plot.
	In each panel, the dashed curves indicate the analytic predictions of DST, while the symbols present numerical estimates based upon MC simulations and MBAR calculations.
	The error bars indicate the estimated uncertainty in these numerical calculations based upon the standard deviation in the results of three independent MC simulations.
%	Systems with a turnover in their interfacial properties.
%	(a) Preferential interaction coefficients as a function of \osm{} \conc{}.
%	The purple and red circles in the inset represent two systems with opposite, non-monotonic behavior. Both systems have a small interfacial preference ($k \approx 0$) but different phase separation propensities ($\chi$). Colors on the heatmap represent the curvature of the surface tension, $h$.
%    (b) Surface tension for the two systems as a function of \osm{} \conc{}.
%    Error bars represent the standard deviation across three independent simulations.
}
	\label{fig:turnover}
\end{figure}

DST suggests that intermolecular interactions can convert surfactants into effective depletants and vice versa.
Moreover, Eq.~\eqref{eq:binary_theory_pref1} indicates that, if $k$ and $h$ have the same signs, then $\Ga$ can switch signs at a critical concentration, $x_* = k/h$.
%Since we consider only $x \le 0.05$ and $| h | \le 3$ for the solutes we consider, it follows that $|k| \le 3/20$ in order for us to observe such a transition, i.e., the solute must be nearly surface-neutral.
Since we only consider $x \le 0.05$, it follows that $h/k \ge 20$ for us to observe this transition. 
For the simulated models, we find that this transition is only observed when $k \approx 0$, i.e., the solute must be nearly surface-neutral.

Figure~\ref{fig:turnover} illustrates two possible transitions. 
The purple curve presents results for attractive solutes with a slight preference for the interface ($k \gtrsim 0$) and $h > 0$.
At infinite dilution, these solutes begin to accumulate at the interface, $\left. \dd\Ga/\dd x \right|_0= k > 0$.
However, because $\dd\tk/\dd x = - 2 ( h + k \ove) \approx - 2 h < 0$, the effective interfacial preference decreases with increasing solute concentration.
Consequently, the slope in Fig.~\ref{fig:turnover}a, $\dd \Ga / \dd x = \tk$, decreases with increasing $x$. 
%However, at slightly higher concentrations the solutes begin to preferentially partition into the bulk in order to form more solute-solute contacts. 
The surface excess reaches a maximum near $x \approx 0.02$ where the effective interfacial preference vanishes, $\tk = \dd \Ga / \dd x = 0$. 
At higher solute concentration, the weak surfactant has become an effective depletant $\tk < 0$, such that the surface excess begins to decrease and ultimately vanishes near $x_* \approx 0.05$.
At even higher concentrations, the surface excess becomes negative and continues to decrease.
Simultaneously, the surface tension, $\bdtga$, initially decreases but appears to reach a minimum near $x_* \approx 0.05$.
%Consequently, the 
%Consequently, the surface excess, $\Ga$, initially increases because $k > 0$, reaches a maximum at $x \approx 0.02$ where $\tk = 0$, vanishes near $x_* \approx 0.05$, and becomes negative for $x > x_*$.
%Simultaneously, the surface tension, $\bdtga$, initially decreases but appears to reach a minimum near $x \approx 0.05$.
Thus, in this instance, attractive solute-solute interactions converted a weak surfactant into a weak effective depletant.

Conversely, the red curve in Fig.~\ref{fig:turnover} illustrates the opposite transition. 
In this case, the solute demonstrates a slight intrinsic preference towards hydration ($k \lesssim 0$). 
However, repulsive solute-solute interactions cause the effective interfacial preference to increase with increasing solute concentration.
At sufficiently high concentrations, these repulsive interactions drive solutes from the bulk to the interface.
The surface excess initially decreases and passes through a minimum.
Beyond this minimum, the surface excess begins to increase and the solute behaves as a weak effective surfactant, $\tk = \dd\Ga/\dd x > 0$.
Simultaneously, the surface tension initially increases and appears to approach a maximum. 
In this instance, repulsive solute-solute interactions converted a weak depletant into a weak effective surfactant.

In both cases, DST describes the surface tension $\bga$ with almost quantitative accuracy for $x \le 0.05$.
DST also describes $\Ga$ very accurately for $x \le 0.03$. 
At higher concentrations, DST slightly overestimates $\Ga$ for attractive solutes. 
Nevertheless, and more importantly, the MC simulations substantiate the qualitative transitions in surface activity that were predicted by DST.

%%%	ABOVE IS VARUN'S ORIGINAL

%\newpage
\subsection{Interfacial polarization} 
\label{SubSec-ResultsPolarizn}
\begin{figure}[h]
	\ifshowfigures
	\includegraphics{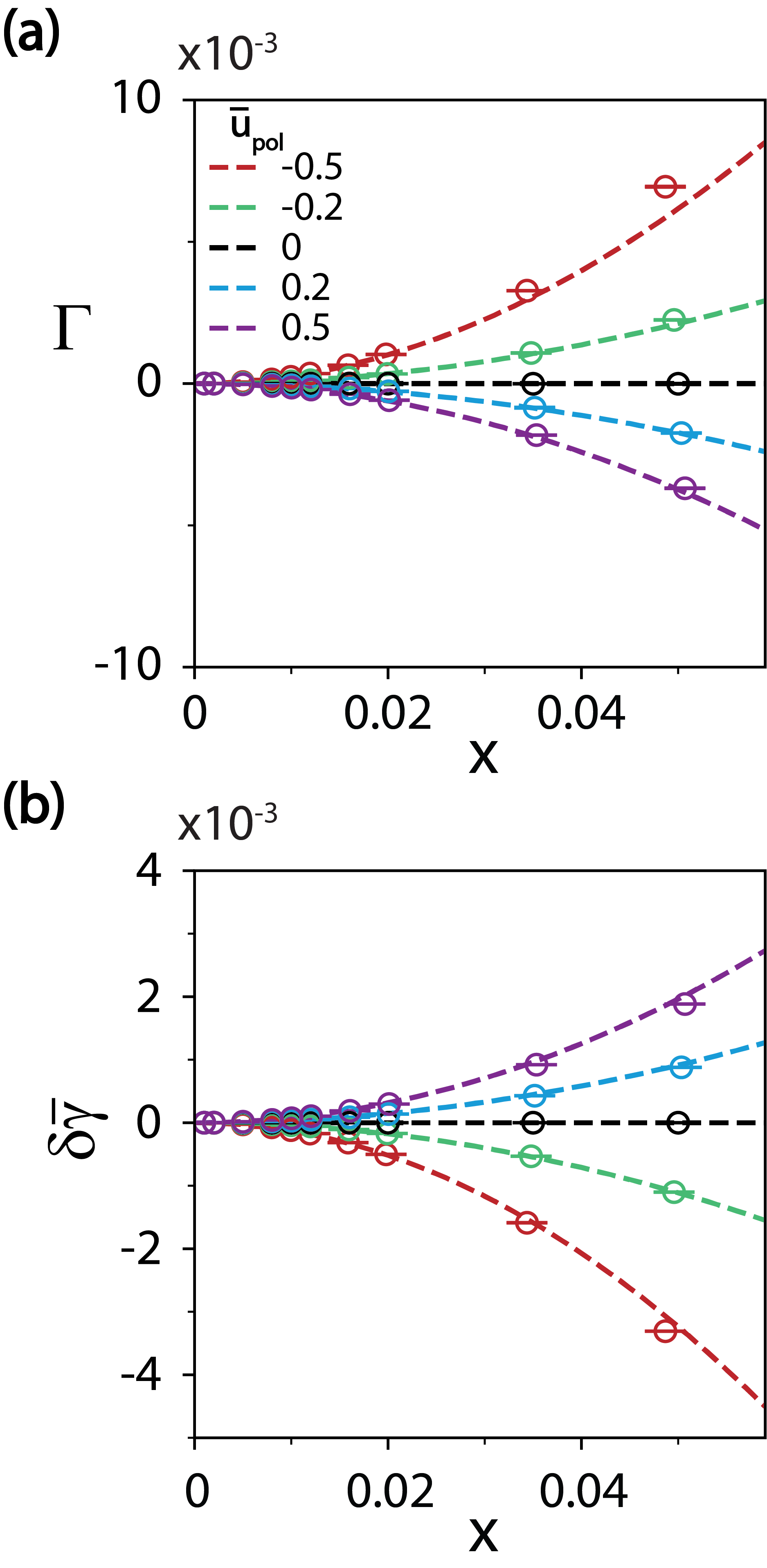}	%Figs/Fig7}
	\fi
	\centering
	\caption{
Influence of polarization upon interfacial properties of ideal solutes ($\Dtbub = 0$) with no intrinsic interfacial preference ($k = 0$). 
Panels a and b present the surface excess, $\Ga$, and surface tension variation, $\dt\bga$, respectively, as a function of solute concentration, $x$.
The dashed curves present the predictions of DST, while the symbols present the results of MC simulations.  
The various colors correspond to the indicated values for the polarization energy ($\bupol $). 
The error bars indicate the estimated uncertainty in the simulation results based upon the standard deviation of three independent MC simulations
%	Effects of incorporating a polarization energy (a) Preferential interaction coefficient and (b) Surface tension as a function of solute concentration for systems with differing polarization energies. 
%Each system has $\chi = 0$ and $k = 0$. The black line represents the case with $u_{pol} = 0$ while the other colors correspond to finite polarization energies.
%Error bars represent the standard deviation across three independent simulations.
}
\label{fig:polarization}
\end{figure}

To this point we have neglected surface polarization effects and assumed that the same contact energies describe intermolecular interactions in both the bulk and interfacial regions of the inhomogeneous system.
In this subsection, we briefly consider the influence of polarization effects in this simple lattice model.
% by including environment-dependent interactions.
The formalism of subsection~\ref{SubSec-StatMech} still applies as before, although now the relevant partition functions become slightly more complex. 
%Moreover, one expects that these polarization effects should have no effect upon either the bulk thermodynamic properties or the intrinsic interfacial preferences of solutes, since the latter can be treated by the solute-interfacial energy parameter. 
%However, environment-dependent interactions will alter the two-solute partition functions and, thus, the curvature, $h$, of the surface tension.

As a first investigation into these polarization effects, we assume that the interface alters only the solute-solute interaction.
Specifically, we assume that pairs of contacting solutes interact with a modified energy $\buoo + \bupol$ when both are at the interface.
Pairs of contacting solutes still interact with an energy $\buoo$ when one or both solutes are in the bulk region.
Consequently, this polarization energy only enters into DST through the 2-solute partition function that determines $h$.
The SM explicitly describes these modifications.

For simplicity, we focus on the special case of ideal solutes ($\Dtbub = 0$) with no surface preference ($k = 0$).
In the absence of polarization, DST predicts that $h = 0$ for ideal solutes with no surface preference.
In this case, DST predicts that $\Ga = 0$ and $\ga = \ga^\phi$ for all solute concentrations.
Conversely, DST predicts that a finite polarization energy, $\bupol \ne 0 $, introduces a finite curvature parameter, $h = h_{\rm pol} \ne 0$, such that
$\Ga = - h_{\rm pol}  x^2 + \mO(x^3)$ and $\dt\bga = \half h_{\rm pol} x^2 + \mO(x^3)$.

Figure~\ref{fig:polarization} assesses these predictions for ideal solutes with no intrinsic interfacial preference.
% the influence of solute polarization upon the surface excess and surface tension for these ideal solutes. 
The black curves indicate that, as expected, in the absence of polarization $\Ga = 0$ and $\ga = \ga^\phi$ for all solute concentrations.
When the polarization energy is attractive ($\bupol < 0$), then $h_{\rm pol} < 0$ such that $\Ga$ increases and $\ga$ decreases quadratically with solute concentration.
Conversely, when the polarization energy is repulsive ($\bupol > 0$), then $h_{\rm pol} >0$ such that $\Ga$ decreases and $\ga$ increases quadratically with solute concentration.
As might be expected, the surface polarization generates a larger effect when $\bupol$ is attractive.
At the highest concentrations $x \ge 0.03$, DST slightly underestimates the simulated effects for the strongest polarization attraction with $\bupol = -0.5$.
Otherwise, DST describes $\Ga$ and $\ga$ with nearly quantitative accuracy for $x \le 0.05$.

%% ABOVE IS VARUN"S ORIGINAL

%\newpage

\subsection{Experimental analysis of osmolyte-water solutions}
\begin{figure}[h]
	\ifshowfigures
	\includegraphics[scale=0.8]{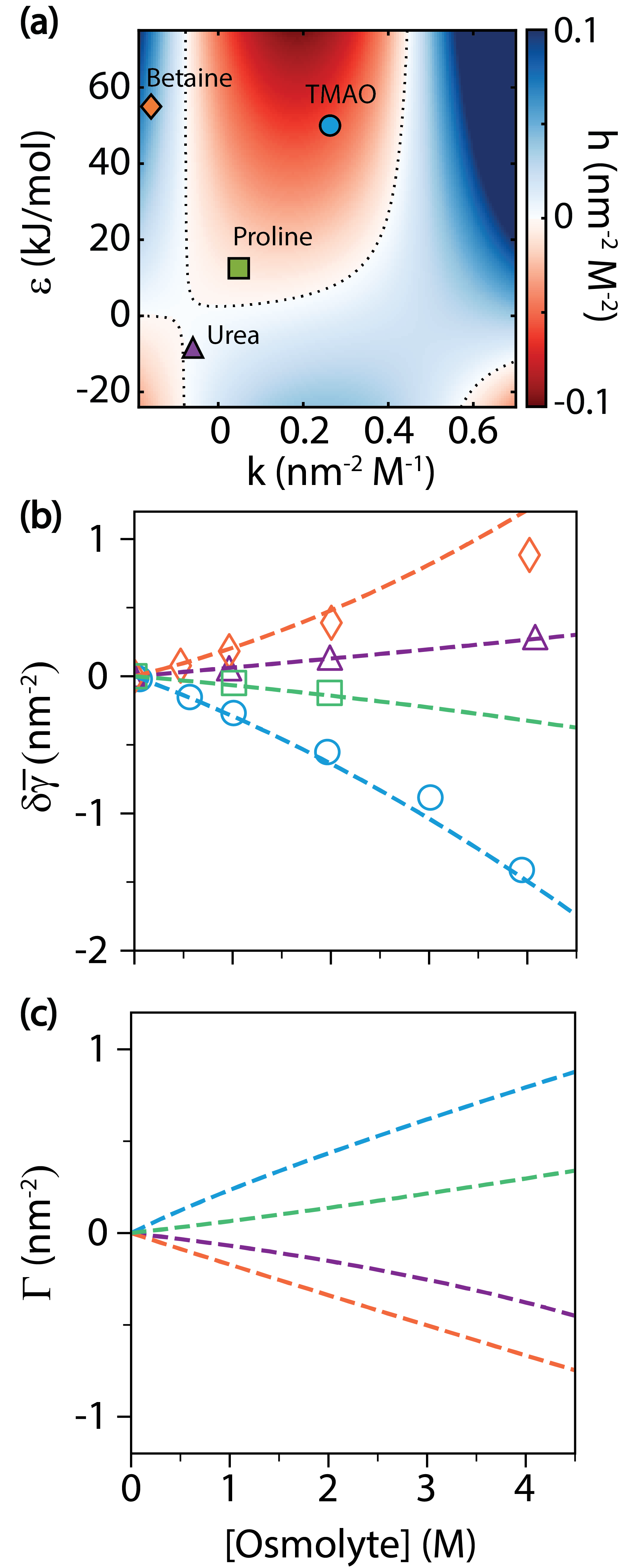}	%Figs/Fig8}
	\fi
	\centering
	\caption{
	DST interpretation of experimental measurements for bulk and interfacial properties of common osmolytes.
(a) 	 Predicted intensity plot for $h$ as a function of the intrinsic interfacial preference, $k$, and bulk energetic, $\ve$, parameters determined from experimental measurements. 
The symbols indicate experimental measurements for betaine, urea, proline, and TMAO.
(b)	Comparison of experimental measurements (symbols) and DST predictions (dashed curves) for the surface tension, $\dt\bga$, of each solute. 
(c) 	DST predictions for the surface excess, $\Ga$, of each solute.
Orange, purple, green, and blue colors correspond to results for betaine, urea, proline, and TMAO, respectively.
%	Understanding real osmolyte solutions via our formalism.
%	(a)  Different colored symbols represent four different solutes, mapped according to their interfacial preference ($k$) and their bulk interaction parameter $\ve$. Colors on the heatmap represent the curvature of the surface tension, $h$.
%	(b) Surface tension for the solutes presented in (a). Data for betaine, urea, proline taken from Auton et. al. \cite{auton2006} and data for TMAO taken from Liao et. al. \cite{liao2017}
%	(c) Predicted preferential interaction coefficients as a function of concentration for the four solutes presented in (a).
}
	\label{fig:experiment}
\end{figure}

Finally, we briefly employ DST to interpret experimental measurements for osmolytes that are commonly employed to modulate protein stability.
Figure~\ref{fig:heatmaps_surften}c demonstrates that,  for the simple nearest-neighbor cubic lattice model without polarization effects, the curvature parameter, $h$, can be determined from the intrinsic interfacial preference, $k$, and the bulk interaction parameter, $\ove$.
Experimental measurements of the surface tension for infinitely dilute solutions allow one to directly determine $k$.\cite{auton2006} 
Moreover, $\ove$ can be directly determined from vapor pressure osmometry (VPO) experiments.\cite{courtenay2000}
However, while the simple cubic lattice model requires that $\ve \le (\zb + 1) \kT \approx$~17 kJ/mol at room temperature, VPO experiments indicate that $\ve \ge $ 50~kJ/mol for betaine.
We employ a simplistic approach to adapt the lattice model for such large effective repulsions by increasing the bulk coordination number from $\zb = 6$ to $\hat z_b = 40$.
Given such a large interaction range, we assume that molecules can retain only $d_i = \hat z_i / \hat z_b = \half$ of their interactions at the interface.
Accordingly, we increase the interfacial coordination number from $z_i = 5$ to $\hat z_i = d_i \hat z_b = 20$.
Given these lattice coordination numbers, we determine the microscopic bulk interaction parameter, $\Dtbub$, from $\ve$ and the interfacial energy parameter, $\Dtbui$, from $k$.
We can then determine $h$ from Eq.~\eqref{eq:binary_beta}.
We observe that $h$ depends quite strongly upon $d_i$, but is otherwise rather insensitive to $\hat z_b$ and $\hat z_i$.
Section~\ref{SubSec-AnalExptlData} and the SM describe our analysis of experimental data in greater detail.

Figure~\ref{fig:experiment}a presents an intensity plot of the calculated $h$ parameter as a function of experimental measurements for $k$ and $\ve$.
The symbols indicate experimental measurements taken from Ref.~\onlinecite{courtenay2000} and \onlinecite{auton2006} for four common osmolytes.
This intensity plot clearly indicates the various regions discussed in Section~\ref{SubSec-ResultsInhomogIntro}.
On this plot, betaine appears to be a strong surfactant with strongly repulsive interactions.
%Consequently, the model predicts that the surface tension will not only increase with betaine concentration but also demonstrate positive curvature, although the surface excess will decrease linearly with concentration.
Conversely, urea, proline, and TMAO all appear in the region characteristic of surface-neutral molecules. 
%In particular, because urea is a weak depletant with slightly attractive interactions, the model predicts that it will become increasingly depleted with   the surface tension will vary linearly with concentration 
In particular, urea appears to be a weak depletant with slightly attractive interactions, proline appears to be a weak surfactant with slightly repulsive interactions, and TMAO appears to be a slightly stronger surfactant with considerably stronger repulsive interactions.
As prior studies have emphasized,\cite{auton2006} it is rather curious that urea denatures folded proteins but is depleted from the air-water interface, while TMAO stabilizes folded proteins but accumulates at the air-water interface.

Given $k$, $\ve$, and $h$, the curves in Fig.~\ref{fig:experiment}b  present the predictions of DST for the surface tension variation, $\bdtga$, of each solute.
The symbols in Fig.~\ref{fig:experiment}b present experimental measurements of the surface tension from Ref.~\onlinecite{auton2006}.
The model slightly overestimates the positive curvature of the measured surface tension for betaine.
Otherwise, the model describes the experimental measurements with rather striking accuracy.
In particular, the model accurately predicts that the urea surface tension varies linearly with concentration, that the surface tension of proline solutions demonstrates slightly negative curvature, and that the surface tension of TMAO solutions demonstrates even more pronounced negative curvature.

Finally, Fig.~\ref{fig:experiment}c presents the corresponding predictions of DST for the surface excess, $\Ga$, of each solute.
These curves illustrate the trends described in Subsection~\ref{SubSec-ResultsInhomogIntro}.
In particular, because betaine is a very repulsive and strongly hydrated solute, its preference for hydration is insensitive to the presence of other betaine molecules (i.e., $\tk(x) \approx k$).
%Consequently, the orange curve in Fig.~\ref{fig:experiment}c varies quite linearly with concentration.
The surface excess also varies approximately linear with concentration for proline because it is a weakly repulsive solute with a slight preference for the interface such that $h + k\ove \approx 0$.
Conversely, because urea is a weakly attractive solute with a slight preference for hydration, the presence of attractive interactions with other urea molecules in the bulk enhances its preference for hydration (i.e., $\dd \tk /\dd x < 0$), as indicated by the negative curvature of the purple curve.
Finally, because TMAO is a highly repulsive solute with a slight preference for the interface, the negative curvature of the blue curve indicates that, as the concentration of TMAO increases, it becomes a less effective surfactant (i.e., $\dd\tk/\dd x < 0$) due to repulsive interactions with other TMAO molecules at the interface.

%%% ABOVE HERE IS VARUN'S ORIGINAL

%\newpage

\section{Conclusions}
In this work we have investigated the influence of solute-solute interactions upon the interfacial properties of multicomponent solutions. 
We first presented a simple thermodynamic model for the surface excess of $o$-solutes, $\Gao$, based upon bulk and interfacial properties that can be experimentally measured.
This model reflects the sensitivity of the surface tension, $b_o = \partial (\bt \ga)/\partial m_o$, and the chemical potential for $o'$-solutes, $G_{o'o} = \partial (\bt \mu_\op)/\partial \mo$, to the molality of $o$-solutes.  
%Specifically, this model depends upon both the slope of the surface tension, $b_o = \partial (\bt \ga)/\partial m_o$, with respect to the (dimensionless) molality, $m_o$, of the bulk solution, as well as the the sensitivity of the chemical potential, $\muo$, for $o$-solutes to the molality of $o'$-solutes, $G_\oop = \partial (\bt \muo)/\partial \mu_\oop$.
%experimental measurements of the surface tension and bulk interactions.
%In particular, experimental measurements determine the slope of the surface tension, $b_o = \partial (\bt \ga)/\partial m_o$, with respect to the (dimensionless) molality, $m_o$, of the bulk solution. 
%Moreover, VPO experiments can determine the sensitivity of the chemical potential, $\muo$, for $o$-solutes to the molality of $o'$-solutes, $G_\oop = \partial (\bt \muo)/\partial \mu_\oop$.
The Gibbs adsorption equation then implies $\Ga = - G\inv b$, which should apply quite generally for multicomponent and concentrated solutions.

In particular, this general result simplifies for dilute binary solutions to 
\begin{equation*}
\Ga = m \left( \frac{k-hm}{1 +  \ove m}\right) 	,
\end{equation*}
where  $k = \left. \partial (\bt \ga)/\partial m\right|_0$ and $h = \left. \partial^2 (\bt \ga)/\partial m^2\right|_0$ describe the slope and curvature of the surface tension with respect to solute molality, respectively, while  $\ove = \left. G \right|_0 - m\inv $ describes bulk interactions between solutes. 
This result clearly indicates that intermolecular interactions can modulate the surface activity of dilute solutes. 
Moreover, it suggests an important distinction between the ``intrinsic'' interfacial preference of solutes, $k = \left. \partial \Ga / \partial m\right|_0$, and the ``effective'' interfacial preference of solutes, $\tk(m) = \partial \Ga / \partial m = k - 2 ( h + k \ove) m$.
The intrinsic preference, $k$, depends only upon the properties of a single solute molecule in pure solvent.
In contrast, the effective preference, $\tk$, depends upon the solution composition and reflects interactions with solutes that are already present in the bulk and at the interface.
%Specifically, the interactions with the solutes that are already present in the bulk and at the interface can clearly modulate the tendency of additional solutes for partitioning between these two regions. 

We next developed a statistical mechanics framework for determining the thermodynamic parameters, $k$, $h$, and $\ove$. 
In the present work, we adopted a simple lattice model for inhomogeneous solutions due to its computational and conceptual advantages. 
This model provides a particularly direct and transparent link between thermodynamic parameters and intermolecular interactions, although it does not properly describe molecular shape, packing, or conformational freedom. 
Importantly, though, the statistical mechanical framework is rather general and should be readily extended for more realistic off-lattice models, albeit at the cost of increased computational cost and conceptual complexity.

Following Hill's earlier work,\cite{hill_theorysolns_1957} we developed a dilute solution theory (DST) for homogeneous solutions that provides the lowest order approximation for treating solute-solute interactions.
The resulting expression for the solute chemical potential has the same form as regular solution theory (RST).
However, in contrast to RST and other mean field theories that are often adopted for modeling concentrated solutions,\cite{hill1986,rubinstein2003} we did not adopt a random mixing approximation. 
In particular, DST predicts that $\ove$ depends exponentially, rather than linearly, upon the microscopic solute-solute dimerization energy, $\Dtbub$. 
Our simulations demonstrate that  DST provides a much more accurate description of solute-solute interactions in dilute solutions.
Moreover, although DST cannot describe concentrated solutions or liquid-liquid phase separation, it appears reasonably accurate up to molar concentrations for the model systems that we consider.

We then extended this formalism to determine the influence of dilute solutes upon interfacial properties. 
We derived the surface tension and surface excess from an analogous perturbation theory in the bulk solute activity by relating the partition function of an open inhomogeneous system to the partition function for a coexisting bulk solution. 
%The seminal study of Guggenheim and other approaches that treat the interface as a semi-autonomous phase and often adopt a random mixing mixing approximation, we derived the surface tension from an analogous perturbation theory that related the partition functions of coexisting homogeneous and inhomogeneous systems. 
The intrinsic interfacial preference, $k$, is related to the ratio of effective partition functions for an infinitely dilute solute at the interface and in the bulk region. 
This intrinsic interfacial preference then depends only upon the microscopic interfacial energy, $\Dtbui$.
In the absence of surface polarization effects, the curvature of the surface tension, $h$, is determined by the microscopic bulk and interfacial energies, $\Dtbub$ and $\Dtbui$.
Thus, in this case, $h$ can be inferred from experimental measurements of $\ove$ and $k$. 
Intuitively, one might imagine that the surface tension should vary linearly with solute concentration (i.e., $h = 0$) only for weakly interacting solutes.
However, our study suggests that $h$ can vanish for a wide range of solutes, including those with strong intermolecular interactions.

For this simple lattice model, the intrinsic interfacial preference, $k$, can be used to distinguish three classes of solutes. 
For strong depletants ($k \lesssim -0.5$), $\ove$ and $h$ tend to have similar values such that $\dd\tk/\dd x \approx - 2(h+k\ove) \approx 0$.
Consequently, solute-solute interactions have relatively little influence upon the surface preferences of strong depletants. 
For strong surfactants ($k \gtrsim 1$), $\ove$ and $h$ again tend to have the same sign. 
In this case, though, solute-solute interactions more significantly impact interfacial activity.
Specifically, repulsive interactions reduce the effective interfacial preference of surfactants, while attractive interactions enhance this effective preference.
Thus, interactions at the interface appear to dominate the surface activity of strong surfactants.

Importantly, our study also suggests a third class of surface neutral solutes  ($-0.5 \lesssim k \lesssim 1$) for which $\ove$ and $h$ tend to have opposite sign. 
In contrast to strong surfactants, bulk interactions appear to dominate the surface activity of these solutes. 
For this class, repulsive interactions tend to drive surface neutral molecules to the interface, while attractive interactions tend to drive surface neutral molecules into the bulk. 
Moreover, in this case, solute-solute interactions can overwhelm relatively weak intrinsic preferences for the interface.
%Indeed the numerical simulations in Fig.~\ref{fig:turnover} confirm this predicted transition.
Specifically, repulsive interactions can convert a weak depletant into a weak effective surfactant.
Conversely, attractive interactions can convert a weak surfactant into a weak effective depletant.

Numerical simulations generally confirm the predictions of DST for the class of lattice models that we consider. 
The largest discrepancies arise when attractive surfactants begin to aggregate. 
Nevertheless, even in this case, DST reasonably predicts both the surface excess and surface tension increment up to $x \approx 0.02$, which corresponds to approximately a 1~M solution in water. 
Furthermore, MC simulations demonstrate that this theory can treat surface polarization effects.

We na\"ively extended the present model to treat experimental data for several solutes that are commonly employed to modulate the thermodynamic stability of proteins.
Our model classifies betaine as a strongly depletant, while urea, proline, and TMAO all appear to be surface neutral molecules.
The extended model describes experimental measurements of the surface tension with surprising accuracy and also predicts the surface excess of each solute.

The present work indicates several future directions.
In particular, future studies should consider solutions with multiple distinct cosolute species.
We anticipate that the cross-interaction between distinct cosolute species may dramatically alter interfacial preferences. 
It will also  be important to extend this framework for more realistic off-lattice models.
Furthermore, it may be possible to extend this framework to more concentrated solutions in analogy to Kirkwood-Buff theory.\cite{KIRKWOOD:1951fr,smith2006}
Moreover, we anticipate extending this framework to investigate non-additive effects of cosolvent mixtures upon the thermodynamic stability of proteins and other macromolecules.\cite{rosgen_molecular_2007,rosgen_synergy_2015} 
Nevertheless, we hope that this first study may provide useful insight into the influence of intermolecular interactions upon interfaces.

\begin{comment}
This paper presents a new theory to describe the surface tension of mixtures based on a semi-grand canonical description of a lattice. We derive a general framework for an arbitrary number of solutes but specialize the discussion to binary systems in this study. A future study will consider ternary solutions and focus on the impact of the cross interaction on interfacial properties.

We showed that our theory predicts bulk free energies and outperforms mean field theory.

Systems with surface active solutes present interfacial phenomena that are strongly correlated with the solute's interfacial preference. Non-surface active solutes present more interesting (albeit smaller) changes to the surface tension. Depending on the balance of the \osm{}-\osm{} interactions and the interfacial preference, turnover behaviors may also be observed.

We also account for the presence of any potential polarization due to the interface and show that these effects can be understood within our framework.

Finally, we show that our formalism may be applied to experimental systems. Given low \conc{} surface tension data and VPO data, our theory predicts the curvature of the surface tension as well as preferential interaction coefficients.
\end{comment}

%\newpage
\section{Methods}
This section describes the numerical calculations presented in this work.
The SM provides additional information about these computational methods.
\label{Sec-Methods}

\subsection{Microscopic lattice model}
We consider a microscopic lattice model for binary solutions at a fixed temperature, $T$.
We describe the system by a $D = 3$ dimensional cubic lattice of $M = L^D$ sites with a coordination number $\zb = 2D = 6 $.
Unless otherwise specified, $L = 10$ such that $M = 10^3$.
Each site is occupied by either a water, $w$, or a solute, $o$, molecule.
The configuration, $\bR$, of the system is specified by the location of the solute molecules.

%\subsubsection{Buk model}
We model a homogeneous bulk fluid with $\no$ solutes by employing periodic boundary conditions in all three directions.
We assume a nearest-neighbor contact potential that is specified by the water-water ($\uww$), solute-water ($\uow$), and solute-solute ($\uoo$) interaction parameters:
\begin{equation}
\label{def-nnlaticeU}
U(\bR;\no) 
	= 
		\sum_{\llangle i, j \rrangle} u_{t_i t_j} 
	= 
		\Unull 
%	+ 	\no \zb \left( \uow - \uww \right) 
	+ 	\no \zb \dt \uow 
%	+ 	\noo(\bR) \left( \uoo + \uww - 2\uow \right) 
	+ 	\noo(\bR) \Dt u_b 
	.
\end{equation}
Here the summation is over all pairs of neighboring sites on the periodic lattice, 
$t_i$ specifies the type of molecule occupying site $i$, 
$\Unull = \half \zb \uww$ is the potential for pure solvent, 
$\dt\uow = \uow - \uww$ is the difference between $o$-$w$ and $w$-$w$ contact energies, 
$\Dt u_b = \uoo + \uww - 2\uow$ is the energy of solute dimerization, 
and 
$\noo(\bR)$ is the number of solute-solute contacts in configuration $\bR$.
We fix the energy scale of the model by defining $\buww = \bt \uww = -1.$

We model an inhomogeneous system of $\No$ solutes by employing periodic boundary conditions in the x and y- directions, but not in the z-direction.
Moreover, we assume that the set $\cS$ of $M_i = 2 L^2$ sites in the top ($z=1$) and bottom ($z=L$) layers of the lattice experience an external field.
We mimic a liquid-vapor interface by assuming this external field vanishes for water molecules, i.e., $\bt\uwv = 0$.
We vary the interfacial preference of solutes by varying $\bt\uov$.
The potential for this inhomogeneous system is  
\begin{eqnarray}
\label{def-Ut}
U_t(\bR;\No) 
	& = & 
		\sum_{\llangle i, j \rrangle} u_{t_i t_j} + \sum_{i\in\cS} u_{t_i v}		\\
	& = & 
		\Unull 
%		+ \No \zb \left( \uow - \uww \right) 
		+ \No \zb \dt\uow 
%		+ \noo(\bR) \left( \uoo + \uww - 2\uow \right) 	\nonumber\\
		+ \noo(\bR) \Dt u_b 		%\nonumber\\
%	&& 
%		+ M_i \left(\uwx - \uww\right) 
		+ M_i \dt \uwv %\left(\uwx - \uww\right) 
%		+ N_{oi}(\bR) \left( \uox - \uwx \right)	, 
		+ N_{o|i}(\bR) \dt\uov %\left( \uox - \uwx \right)	, 
		,
\end{eqnarray}
where 
the sum over $i\in\cS$ denotes a sum over the $M_i$ interfacial sites,
$\dt\uwv = \uwv - \uww$, 
$\dt\uov = \uov - \uwv$, 
and
$N_{o|i}(\bR)$ denotes the number of solutes at the interface in configuration $\bR$.
%The sum over $i\in\cS$ denotes a sum over the $M_i$ interfacial sites, 
The term 
%$M_i \left( \uwx - \uww \right)$ 
$M_i \dt\uwv$ 
accounts for the loss of water-water contacts at the interface in the case of pure solvent, 
%$N_{ox}(\bR)$ denotes the number of solutes at the interface, 
and 
%$N_{ox}(\bR) \left( \uox - \uwx \right)$
$N_{o|i}(\bR) \dt\uov$
 is an energetic cost associated with interfacial solutes.
%We specialize to the liquid-vapor interface by assuming that $\bt\uwx = 0$.
%We use $\uox = \uox - \uwx$ to describe the relative preference of the solute for the interface.
Note that Eq.~\eqref{def-Ut} assumes the same contact energies in the bulk and interfacial regions.
In the case that polarization effects are included, Eq.~\eqref{def-Ut} is supplemented with an additional energy $u_{\rm pol}$ for each pair of contacting solutes that are both at the interface.

\subsection{Monte Carlo simulations}
We identified each site of the lattice by x, y, and z-coordinates that corresponded to a unique integer between 1 and $M$. 
In the case of the inhomogeneous lattice, the first and last $L^2$ integers corresponded to the top and bottom layers of the lattice. 
We simulated the lattice model with a simple canonical Monte Carlo algorithm with fixed composition.
We represented each configuration, $\bR$, by a permutation, $p$, of the integers $1, \ldots, M$, in which the first elements of the permutation indicate the lattice sites occupied by solutes.
We generated a new trial configuration, $\bR'$, by employing the Fisher-Yates shuffling algorithm\cite{knuth1997} to create a new permutation, $p'$.
Consequently, all trial configurations, $\bR'$, are generated with equal probability.
We then employed the standard Metropolis criterion to accept or reject the trial move:
\begin{equation}
\text{Acc}(\bR\to\bR') = \min\{ 1, \exp\left[-\bt\left( U(\bR') - U(\bR) \right) \right]\} 	.
\end{equation}
We tested our MC simulations by comparing the free energies determined from MC simulation with free energies determined via analytic calculation or exhaustive enumeration for systems of tractable complexity. 
While more sophisticated Monte Carlo methods could be adopted, this simple approach proved adequate for the present study.

The MC simulations were performed for $5 \times 10^5$ steps for bulk systems and $1 \times 10^6$ steps for inhomogenous systems. The first $10\%$ of steps were discarded for equilibration purposes.
Statistics were sampled from the remainder of the simulation every 10 steps.

%\Varun{We need to add information about the following:}
%The MC simulations were performed for XX steps. 
%The first YY steps were discarded for equilibration purposes.
%Energies and ZZ were then sampled every AA steps.

%\Varun{We should say something about how the sites/locations on the lattice were encoded, where the interface was, and how you checked for contacts.}

\newcommand{\dAdst}{\dt A_{\rm DST}}

\subsection{Free energy calculations}
\subsubsection{Homogeneous bulk system}
We consider a solvent and solute described by the interaction parameters $u = (\uww,\uow,\uoo)$.
We are interested in the free energy change, $\dA(\no;u) = A(\no;u) - A(0;u) = M \{ \dtmuw(\xo;u) + \xo \muo(\xo;u) \}$, due to introducing $\no$ solutes into a homogeneous bulk solution, where $\xo = \no/M$ and $\dtmuw(\xo;u) = \muw(\xo;u) - \muwo(\uww)$.
DST indicates that this may be approximated
\begin{eqnarray}
\label{def-dAthry}
M\inv \bt\dAdst(\no;u) 
%	& = &  
%		\bt \dtmuw(\xo;u) + \xo \bt\muo(\xo;u) 	\\
\label{eq-dAthry}
	& = &  
%	\kT 	\left(
		\xo \ln \xo - \xo 
%		\right)
	+ 	\bmuost \xo
	+ 	\half \bveoo \xo^2
	+ 	\mO(\xo^3)	, 
\end{eqnarray}
where 
%$\xo = \no/M$ and $\dtmuw(\xo;u) = \muw(\xo;u) - \muwo(\uww)$, while 
$\bmuost = \bt\muost(u)$ and $\bveoo = \bt\veoo(u)$ are determined by Eqs.~\eqref{def-muost} and \eqref{def-veoo}, respectively.
We wish to compare this approximate analytic expression with numerically exact calculations of the free energy difference, which may be expressed:
\begin{equation}
\label{eq-dA-decomp}
\dAnum(\no;u) = \dAcomb(\no) + \dAint(\no;u)	.
\end{equation}
Here the combinatoric term is defined
\begin{equation}
\label{def-dAcomb}
\dAcomb(\no) \equiv -\kT \ln \left[\frac{M!}{\nw!\no!}\right]	,
\end{equation}
where $\nw = M - \no$ is the number of water molecules, while the interaction term is defined
\begin{equation}
\label{def-dAint}
\dAint(\no;u) \equiv -\kT \ln \left[ \QD(\no;u) / \QD(0;u) \right]		,
\end{equation}
where $\QD(\no;u)$ is the canonical partition function for a system of $\no$ solute molecules with interaction parameters, $u$, in the case that all molecules are distinguishable (D).
While $\dAcomb(\no)$ describes the ideal mixing entropy due to the combinatorics of indistinguishable molecules, $\dAint$ quantifies the cost of replacing $w-w$ interactions with $o-w$ and $o-o$ interactions.
Because $\dAint$ is expressed as a ratio of partition functions for indistinguishable molecules, it can be directly estimated from conventional free energy methods.
In the thermodynamic limit for dilute solutions, the combinatoric term simplifies
\begin{eqnarray}
\label{eq-dAcomb-ideal}
M\inv \bt \dAcomb(\no)
	& \stackrel{\no,\nw\gg1}{\longrightarrow}	& 
%	M \kT 
%	\left\{ 
		\xo \ln \xo + \xw \ln \xw
%	\right\}		
						\\
\label{eq-dAcomb-idealdilute}
	&&
	\stackrel{\xo\ll1}{\longrightarrow}
%	M \kT 
%	\left\{ 
		\xo \ln \xo - \xo + \half \xo^2 + \mO(\xo^3)
%	\right\}			.
	.
\end{eqnarray}	
The first expression results from applying Stirling's approximation when $\no,\nw \gg 1$ and corresponds to the ideal entropy of mixing for ideal solutions, while the second expression results from expanding $\ln (1-\xo) = - \xo - \half \xo^2 + \mO(\xo^3)$.
In particular, the $\half\xo^2$ term in Eq.~\eqref{eq-dAcomb-idealdilute} explicitly accounts for the excluded volume interaction that is implicit in $\dAint$.
Thus, in the thermodynamic limit for dilute solutions, the exact free energy difference may be expressed:
\begin{equation}
\label{eq-dAnum}
M\inv\bt\dAnum(\no;u)
%	= 
	\stackrel{\nw \gg \no \gg 1}{\longrightarrow}
%	\kT \left( 
	\xo \ln \xo - \xo 
%	\right) 
	+ \half \xo^2 
	+ M\inv\bt\dAint(\no;u) + \mO(\xo^3) 		.
\end{equation}
Since the first two entropic terms arise in both Eq.~\eqref{eq-dAthry} and \eqref{eq-dAnum}, a critical assessment of the approximate DST should compare the third and fourth terms that describe interactions:
\begin{equation}\label{eq:bulk_free_energy_diff}
\bmuost \xo + \half \bveoo \xo^2 		
	\stackrel{?}{=} 
\half \xo^2 + M\inv\bt\dAint(\no;u)  		
	.
\end{equation}

However, Eq.~\eqref{eq-dAcomb-idealdilute} does not hold for the relatively small systems that we simulate in this study.
This introduces a systematic discrepancy between the analytic approximation and numerically exact calculations of the free energy difference, $\dA(\no;u)$.
This discrepancy is entirely due to finite size errors in treating the combinatoric term, $\dAcomb(\no)$, and is independent of interaction parameters, $u$.
Consequently, we introduce a reference system with interaction parameters, $u_r$,  and define 
\begin{equation}
\label{def-Dtba}
\Dt \ba(\no;u) \equiv M\inv \bt \left\{ \dA(\no;u) - \dA(\no;u_r) \right\}	.
\end{equation}
%to cancel out this finite size discrepancy, which contributes equally to $\dA(\no;u)$ and $\dA(\no;u_r)$.
Because $\dAcomb(\no)$ contributes equally to $\dA(\no;u)$ and $\dA(\no;u_r)$, the difference $\Dt \ba(\no;u)$ cancels out this finite size discrepancy.
For convenience, we define a reference state in which the solutes interact identically to solvent molecules, i.e., $u_r = (\uww,\uww,\uww)$.
For this reference state, $\bt\dAint(\no;u_r) = 0$ for all $\no$, while $\bmu_{o;r}^* = \bt\muost(u_r) = 0$ and $\ove_{oo;r} = \bt\ve_{oo}(u_r) = 1$.
\begin{comment}
Given this reference state:
\begin{equation}
\Dt a(\no;u)
=
	\begin{cases}
		M\inv \bt \dAint(\no;u) 					& \text{from numerical calculations}	\\
		\bmuost \xo + \half \left(\bveoo - 1\right) \xo^2  	& \text{from analytical theory}
	\end{cases}
\end{equation}
\end{comment}

While $\Dt \ba$ properly eliminates the finite size artifacts, it is often dominated by the linear term $\bmuost \xo$ in Eq.~\eqref{eq-dAthry} that corresponds to solutes at infinite dilution. 
Consequently, in order to critically assess the accuracy of DST for modeling solute-solute interactions, we subtract this infinite dilution contribution and define the (scaled) interaction free energy
\begin{eqnarray}
\Dt \baint(\no;u) 
	& \equiv & 
		\Dt \ba(\no;u) - \bmuost \xo		\\
	& = & 
\label{eq-Dtbaint}
	\begin{cases}
		\half \left(\bveoo - 1\right) \xo^2  	& \text{DST}							\\
		M\inv \bt \dAint(\no;u) - \xo \bmuost	& \text{numerically exact calculations}	
	\end{cases}
\end{eqnarray}
We employed the Python implementation of the multistate Bennett acceptance ratio (MBAR) method\cite{shirts2008} to estimate $\dAint(\no;u)$ from Eq.~\eqref{def-dAint} based upon configurations sampled from MC simulations with $\no$ solutes interacting according to the parameters $u$.

\subsubsection{Inhomogeneous interfacial system}
Similarly, we are are interested in the free energy change, $\dA_t(\No;u_t) = A_t(\No;u_t) - A_t(0;u_t)$, associated with adding $\No$ solutes into an inhomogeneous system described by the interaction parameters $u_t = (u;\uwv,\uov) = (\uww,\uow,\uoo;\uwv,\uov)$.
The same finite size discrepancy in the entropy of mixing also arises in considering the free energy of the inhomogeneous system.
Accordingly, we define an inhomogeneous reference system with interaction parameters, $\utr = (\ur;\uwv,\uwv) = (\uww,\uww,\uww;\uwv,\uwv)$, such that the solutes behave equivalently to solvent molecules.
The system and the reference system have the same chemical potential, $\muwo$, and surface tension, $\ga^\phi$, for pure solvent, such that $A_t(0;\ut) = A_t(0;\utr) = M \muwo + M_i \ga^\phi$.
We then define an analogous dimensionless (scaled) free energy difference:
\begin{eqnarray}
\label{def-Dtat}
\Dt \ba_t(\No;\ut) 
	& \equiv & 
		M\inv \bt 
		\left\{ 
			\dt A_t(\No;\ut) - \dt A_t(\No;\utr) 
		\right\}	
		\\
	& = & 
	- M\inv \ln \left[ Q_{t\rm D}(\No;\ut)/Q_{t\rm D}(0;\ut)\right]	,
\end{eqnarray}
where $Q_{t\rm D}(\No;\ut)$ is the distinguishable partition function for the inhomogeneous system.
The second expression follows because the combinatoric terms cancel between the two systems and because $Q_{t\rm D}(\No;\utr) = Q_{t\rm D}(0;\utr)$ for the reference system.
Moreover, in order to eliminate the contributions from the bulk region, we define 
\begin{equation}
\label{def-Phi}
\Phi(\No;\ut) 
	\equiv \Dt \ba_t(\No;\ut) - \Dt \ba(\No;u)  
	= 
		- M\inv 
		\ln\left[ 
			\left. 
			\frac{Q_{t\rm D}(\No;\ut)}{Q_{t\rm D}(0;\ut)} 
			\right/
			\frac{Q_{\rm D}(\No;u)}{Q_{\rm D}(0;u)} 
		\right]	,
\end{equation}
which is determined from ratios of partition functions that can be readily estimated.

We now wish to compare numerically exact calculations for $\Phi(\No;\ut)$ with the predictions of DST.
However, the presence of the interface introduces additional complications into this comparison.
We define $\xot \equiv \No/M$ as the total fraction of lattice sites in the inhomogeneous system that are occupied by solute molecules.
We then define 
\begin{equation}
\txo = \txo(\No;\ut) \equiv \txo(\No,M,M_i;\ut) 
\end{equation}
as the solute mole fraction in the bulk region of the inhomogeneous system.
In general, $\txo \neq \xot$ because solute and solvent molecules may demonstrate different preferences for the interface. 
%Elsewhere in this manuscript $\xo$ denotes the mole fraction of solutes in the bulk region.
%However, in this subsection, it is useful to define $\xo \equiv \No/M$ as the total fraction of lattice sites in the inhomogeneous system that are occupied by solute molecules.
%In general, the solute and solvent molecules will demonstrate different preferences for the interface.
%Consequently, the composition of the bulk region is not simply $\xo$.
%We denote 
%\begin{equation}
%\txo = \txo(\No;\ut) \equiv \txo(\No,M,M_i;\ut) 
%\end{equation}
%as the solute mole fraction in the bulk region of the inhomogeneous system.
The thermodynamic expression for the free energy of the inhomogeneous system is then
\begin{equation}
\label{eq-Atut}
A_t(\No;\ut) = M \muw(\txo;u) + \No \muo(\txo;u) + M_i\ga(\txo;\ut)	,
\end{equation}
where $\mu(\txo;u)$ denotes the chemical potential for a homogeneous system with solute mole fraction $\txo$ and bulk interaction parameters $u$.

The reference system is defined such that solute and solvent molecules have equivalent surface activity. 
%In contrast, the solute and solvent molecules behave identically in the reference system.
Consequently, the composition of the bulk region of the reference system is simply $\txo(\No;\utr) = \No/M = \xot$.
Moreover, the surface tension of the reference system is independent of $\No$, i.e., $\ga(\xo;\utr) = \gao$ for all $\xo$.
Consequently, the free energy of the reference system is  
\begin{equation}
\label{eq-Atutr}
A_t(\No;\utr) = M \muw(\xot;u_r) + \No \muo(\xot;u_r) + M_i\gao	.
\end{equation}
%However, if the solute and solvent molecules interact differently, then the bulk composition of the inhomogeneous system is not necessarily $\xo \equiv \No/M$.
Given Eqs.~\eqref{eq-Atut} and \eqref{eq-Atutr}, it follows that
\begin{eqnarray}
\Dt \ba_t(\No;\ut) 
	=
	%	\left[ 
			\bmuw(\txo;u) - \bmuw(\xot;\ur) 
	%	\right]
	+ \xot 
	%\bt 
	\left[ 
		\bmuo(\txo;u) - \bmuo(\xot;\ur) 
	\right]
	+ \eti \bdtga(\txo;\ut) ,
\end{eqnarray}
where $\bmu = \bt\mu$ and $\bdtga(\txo;\ut) = \bt \left( \ga(\txo;\ut) - \gao \right)$.
We compare this with $\Dt \ba(\No;u)$ from Eq.~\eqref{def-Dtba} for a homogeneous system in order to identify the surface contributions:
\begin{eqnarray}
\Dt \ba_t(\No;\ut) - \Dt \ba(\No;u) 
%	& = & 
%	- M\inv 
%		\ln\left[ 
%			\frac{Q_{t\rm D}(\No;\ut)}{Q_{t\rm D}(0;\ut)} 
%			\frac{Q_{\rm D}(0;u)}{Q_{\rm D}(\No;u)} 
%		\right]	\\
	& = & 
		%\bt \left[ 
			\bmuw(\txo;u) - \bmuw(\xot;u) 
		%\right]	
										\nonumber \\
	& & 	
	+ \xot 
		%\bt 
		\left[ 
			\bmuo(\txo;u) - \bmuo(\xot;u) 
		\right]
	+ \eti 
		%\bt 
		\bdtga(\txo;\ut) .
\end{eqnarray}
If we employ the analytic expressions for $\muw$ and $\muo$, we can then obtain a useful expression for the surface tension:
\begin{equation}
\eti %\bt 
	\bdtga(\txo;\ut)
	= 
	\Phi(\No;\ut) 
		+ \dtxo - \xot \ln\left(\frac{\txo}{\xot}\right) 
		+ \half \bveoo \dtxo^2	 
\end{equation} 
where $\dtxo = \txo - \xot$ and $\bveoo = \bt \veoo(u)$.
We employed MBAR to estimate $\Phi(\No;\ut)$ from Eq.~\eqref{def-Phi} based upon statistics from MC simulations.
We compared the resulting numerical estimate for $\dt\bga$ with the DST approximation given by Eq.~\eqref{eq:binary_lattice_surften_fit}.

\subsection{Comparison with Regular Solution theory}
Regular solution theory (RST) provides a useful model for mixing thermodynamics.\cite{hill1986,rubinstein2003}
According to RST, the free energy of the lattice model defined by Eq.~\eqref{def-nnlaticeU} is 
given by
\begin{equation}
M\inv\bt\Arst(\no;u) 
	= 
%		M \kT 
%		\left\{ 
		\half \zb\left( \buoo \xo + \buww \xw \right) 
		+ 
		\xo \ln \xo 	+ \xw \ln \xw
		+ 
		\chi \xo \xw
%		\right\}
\end{equation}
where $\chi = \half \zb \left[ 2 \buow - (\buoo + \buww)\right] = -\half \zb \Dtbub$.
We define $\dt\Arst(\no;u) = \Arst(\no;u) - \Arst(0;u)$ as the free energy cost predicted by RST for introducing $\no$ solutes into the system. 
We again consider a reference state with parameters $u_r = (\uww, \uww,\uww)$ such that solvent and solute molecules interact equivalently and $\chi_r = 0$.
Then the difference, $\Dt \ba_\rst (\no;u) = M\inv\bt\{ \dt\Arst(\no;u) - \dt\Arst(\no;u_r) \}$, in free energy for introducing the solutes into the system and into the reference system is
\begin{equation}
\Dt \ba_\rst (\no;u) 
	=
		\bmuost \xo - \chi \xo^2, 
\end{equation}
where RST determines $\bmuost = \zb ( \buow - \buww )$ in agreement with DST.
Eliminating the contribution from solutes at infinite dilution, RST predicts the interaction free energy to be 
\begin{equation}
\label{eq-DtaRstInt}
\Dt \ba_{\text{RST;int}}(\no;u) = - \chi \xo^2	,
\end{equation}
which should be directly compared with Eq.~\eqref{eq-Dtbaint}.

RST provides a qualitatively accurate description for phase coexistence in concentrated solutions.\cite{hill1986,rubinstein2003}
RST predicts that the binodal, $\chi_b(x)$, and spinodal, $\chi_s(x)$, curves are given by 
\begin{eqnarray}
\label{eq-chib}
\chi_b(\xo) 
	& = & 
		\frac{1}{2 \xo - 1} \ln \left( \frac{\xo}{1-\xo}\right)		\\
\label{eq-chis}
\chi_s(\xo) 
	& = & 
		\half \sqbracket{ \frac{1}{\xo} + \frac{1}{1-\xo}}	.
\end{eqnarray}
These equations for the  binodal and spinodal curves determine corresponding dimerization energies $\Dt\bub = - 2 \chi / \zb$, which are indicated in Fig.~\ref{fig:bulk_results}a.

In contrast, DST cannot model concentrated solutions and cannot describe stable two-phase coexistence.
If we denote $\ba(\xo) \equiv M\inv \bt A(\no)$ as the dimensionless, scaled free energy, then DST predicts that 
\begin{equation}
\frac{\ptl^2 \ba}{\ptl \xo^2} = \xo\inv \left( 1 + \ove \xo \right) ,
\end{equation}
where $\ove=\ove(u)$ is given for the nearest-neighbor cubic lattice model by Eq.~\eqref{eq:binary_bulk_epsilon}. 
Consequently, for attractive solutes with $\ove < 0$, this theory predicts that the solution will become unstable when
\begin{equation}
\xo > - \ove\inv, 
\end{equation}
which is indicated as the DST stability curve in Fig.~\ref{fig:bulk_results}a.

\subsection{Analyzing experimental data}
\label{SubSec-AnalExptlData}
In this subsection we connect the present model to experimental measurements of thermodynamic properties.
The present thermodynamic model for interfaces is completely specified by the parameters, $k$ and $h$.
While $k$ can be directly determined from experimental surface tension measurements for dilute solutions, it is more difficult to determine $h$.
%However, $h$ can be inferred from bulk thermodynamic measurements under certain circumstances.

We first specialize to the case of dilute aqueous solutions.
The molality of pure water is $\hat m_w \equiv 55.5$~mol/kg, while under standard conditions the molarity of pure water is $M_w^\phi \approx $ 55.56~M.
The molarity of pure water determines the size of each lattice site $l_1 = (N_A / M_w^\phi )^{-1/3} \approx $ 0.31~nm, where $N_A$ is Avagadro's number.
This then determines the interfacial area of lattice sites $\sigma_1 = l_1^2 = $ 0.096~nm$^2$.
Moreover, for dilute solutions, the mole fraction $x = \no / ( \no + \nw) \approx \no/\nw \equiv m$, where the dimensionless molality is given by $m = \hat m_o / \hat m_w$ .
Thus, the mole fraction is related to the measured molality, $\hat m_o$, and molarity, $M_o$, of solute by $\xo \approx \hat m_o / \hat m_w \approx M_o / M_w^\phi$.
% by $\hat m_o \approx x \hat m_w$ and to the molarity, $M_o$, by $x \approx M_o / M_w^\phi.$

The dimensionless surface tension of the lattice model, $\bga$, is related to the experimental measurements of the surface tension, $\hat\ga$, by $\bga = \sigma_1 \beta \hat\ga$.
The parameter $k$ can be determined from experimental measurements for dilute solutions:
\begin{equation}
k = - \left. \frac{\ptl \bga}{\ptl \xo} \right|_{\xo = 0}
%  = - \frac{\sigma_1}{\kT} 
%  	\left. 
%		 \left. 
%		 	\frac{\ptl \hat{\ga}}{\ptl M_o} 
%		%\right|_{M_o = 0} 
%	\right/ 
%		%\left. 
%		\frac{\dd \xo}{\dd M_o} \right|_{M_o = 0}
%  = - \frac{\bt \sigma_1}{M_w} 
  = -\bt \sigma_1 M_w^\phi 
 % 	\left. 
		 \left. 
		 	\frac{\ptl \hat{\ga}}{\ptl M_o} 
		\right|_{M_o = 0} 
%	\right/ 
		%\left. 
%		\frac{\dd \xo}{\dd M_o} \right|_{M_o = 0}
\end{equation}

In principle, $h$ can also be determined directly from experimental measurements.
However, it is considerably more challenging to quantitatively determine the curvature of the surface tension for dilute solutions.
Nevertheless, $h$ can be estimated with additional information from bulk measurements. 
Assuming that polarization effects can be neglected, Section~\ref{SubSec-ResultsInhomogIntro} indicates that $h = h(\Dtbub,\Dtbui;\zb,z_i)$ where $\zb$ and $z_i$ are coordination numbers for the bulk and interfacial regions, respectively, while $\Dtbub$ describes the energetics of solute association in the bulk and $\Dtbui$ describes the energetics of moving solutes from the bulk to the interface.
According to Eq.~\eqref{eq-k}, $\Dtbui$ can be directly determined from $k$ and, thus, from surface tension measurements for dilute solutions.
Moreover, Eq.~\eqref{eq:binary_bulk_epsilon} relates the bulk thermodynamic parameter, $\ove$, to $\Dtbub$ and $\zb$.
This bulk thermodynamic parameter can itself be determined from vapor pressure osmometry (VPO).
VPO determines the solvent osmolality:
\begin{equation}
\phi 
%\equiv - \frac{\muw - \muwo}{\hat{m} \kT} 	.
	\equiv 
%		- \frac{1}{\mo} \ln a_w 
		- m_o\inv \ln a_w 
	= 
%		\frac{\bt (\muwo - \muw)}{\mo}
		m_o\inv \bt ( \muwo - \muw ) 
\end{equation}
Employing Eq.~\eqref{eq-muw} for $\muw$ and using $x \approx \hat m_o / \hat m_w$ for dilute solutions, one obtains
\begin{equation}
\phi 
	%\approx M_w^\phi \left( 1 + \half \ove M_w^\phi \hat m \right) 
	\approx
		1 + \half \ove \hat m_o / \hat m_w
\end{equation}
Consequently, $\ove$ can be inferred from VPO measurements for dilute solutions according to
\begin{equation}
\ove 
	= 
%		\frac{2}{\left.M_w^\phi\right.^2} \left. \frac{\dd \phi}{\dd \hat m} \right|_{\hat m = 0}		.
		2 \hat m_w \left. \frac{\dd \phi}{\dd \hat m_o} \right|_{\hat m_o = 0}
\end{equation}
According to Eq.~\eqref{eq:binary_bulk_epsilon}, if one knows the lattice coordination number, $\zb$, then $\ove$ determines the microscopic bulk interaction parameter $\Dtbub$.
In particular, for the simple cubic lattice with $\zb = 6$, $\ve \le (\zb + 1) \kT \approx $ 17 kJ/mol at room temperature. 
Because VPO measurements determine $\ove \approx $ 50 kJ/mol and 55 kJ/mol for TMAO and betaine, respectively, we simplistically redefine $\zb \to 40$ and then determine $\Dtbub$ according to Eq.~\eqref{eq:binary_bulk_epsilon}. 
Finally, we determine $z_i$ by assuming that molecules retain a fraction $d_i = z_i/\zb = 0.50$ of their interactions at the interface. 
The inferred values of $z_i$, $\zb$, $\Dtbub$, and $\Dtbui$, then determine the predictions of the microscopic model for $h$ according to Eq.~\eqref{eq:binary_beta}.
The SM demonstrates that $h$ is relatively insensitive to the precise values of $z_i$ and $\zb$, but is more sensitive to the parameter $d_i = z_i/\zb$.

\section*{Supplementary Material}
See the supplementary material for detailed derivations of the relevant partition functions, for additional details of our computational methods, and for additional numerical results.
%}

\begin{acknowledgments}
%\Red{
Parts of this research were conducted with Advanced CyberInfrastructure computational resources provided
by the Institute for CyberScience at the Pennsylvania State University (http://icds.psu.edu). 
In addition, parts of this research were conducted with XSEDE resources awarded by Grant No. TG-CHE170062. This work used the Extreme Science and Engineering Discovery Environment (XSEDE), which is supported by the National Science Foundation (Grant No. ACI-1548562).\cite{xsede}
%}
The authors gratefully acknowledge Dr.~Pho Bui and Prof.~Paul Cremer for many useful conversations regarding this work.
\end{acknowledgments}

\section*{Author Declarations}
\subsection*{Conflict of interest}
The authors have no conflicts to disclose.

\section*{Data Availablility}
The data that support the findings of this study are available from the corresponding author upon reasonable request.

\bibliography{binary_surften}

%\newpage
%\begin{footnotesize}
%\bibliography{binary_surften}
%\end{footnotesize}
\end{document}